\documentclass[superscriptaddress,aps,prl,showkeys,showpacs,twocolumn,10pt]{revtex4}
\usepackage[OT1,T1]{fontenc}
\usepackage{graphicx}
\usepackage{rotating}
\begin{document}

\title{ABC Effect and Resonance Structure in the Double-Pionic Fusion to
  $^3$He}
\date{\today}

\newcommand*{\IKPUU}{Division of Nuclear Physics, Department of Physics and 
 Astronomy, Uppsala University, Box 516, 75120 Uppsala, Sweden}
\newcommand*{\ASWarsN}{Department of Nuclear Physics, National Centre for 
 Nuclear Research, ul.\ Hoza~69, 00-681, Warsaw, Poland}
\newcommand*{\IPJ}{Institute of Physics, Jagiellonian University, ul.\ 
 Reymonta~4, 30-059 Krak\'{o}w, Poland}
\newcommand*{\PITue}{Physikalisches Institut, Eberhard--Karls--Universit\"at 
 T\"ubingen, Auf der Morgenstelle~14, 72076 T\"ubingen, Germany}
\newcommand*{\Kepler}{Kepler Center f\"ur Astro-- und Teilchenphysik, 
 Physikalisches Institut der Universit\"at T\"ubingen, Auf der 
 Morgenstelle~14, 72076 T\"ubingen, Germany}
\newcommand*{\MS}{Institut f\"ur Kernphysik, Westf\"alische 
 Wilhelms--Universit\"at M\"unster, Wilhelm--Klemm--Str.~9, 48149 M\"unster, 
 Germany}
\newcommand*{\ASWarsH}{High Energy Physics Department, National Centre for 
 Nuclear Research, ul.\ Hoza~69, 00-681, Warsaw, Poland}
\newcommand*{\IITB}{Department of Physics, Indian Institute of Technology 
 Bombay, Powai, Mumbai--400076, Maharashtra, India}
\newcommand*{\Budker}{Budker Institute of Nuclear Physics of SB RAS, 
 11~akademika Lavrentieva prospect, Novosibirsk, 630090, Russia}
\newcommand*{\Novosib}{Novosibirsk State University, 2~Pirogova Str., 
 Novosibirsk, 630090, Russia}
\newcommand*{\IKPJ}{Institut f\"ur Kernphysik, Forschungszentrum J\"ulich, 
 52425 J\"ulich, Germany}
\newcommand*{\JCHP}{J\"ulich Center for Hadron Physics, Forschungszentrum 
 J\"ulich, 52425 J\"ulich, Germany}
\newcommand*{\Bochum}{Institut f\"ur Experimentalphysik I, Ruhr--Universit\"at 
 Bochum, Universit\"atsstr.~150, 44780 Bochum, Germany}
\newcommand*{\ZELJ}{Zentralinstitut f\"ur Engineering, Elektronik und 
 Analytik, Forschungszentrum J\"ulich, 52425 J\"ulich, Germany}
\newcommand*{\Erl}{Physikalisches Institut, 
 Friedrich--Alexander--Universit\"at Erlangen--N\"urnberg, 
 Erwin--Rommel-Str.~1, 91058 Erlangen, Germany}
\newcommand*{\ITEP}{Institute for Theoretical and Experimental Physics, State 
 Scientific Center of the Russian Federation, 25~Bolshaya Cheremushkinskaya, 
 Moscow, 117218, Russia}
\newcommand*{\Giess}{II.\ Physikalisches Institut, 
 Justus--Liebig--Universit\"at Gie{\ss}en, Heinrich--Buff--Ring~16, 35392 
 Giessen, Germany}
\newcommand*{\IITI}{Department of Physics, Indian Institute of Technology 
 Indore, Khandwa Road, Indore--452017, Madhya Pradesh, India}
\newcommand*{\HepGat}{High Energy Physics Division, Petersburg Nuclear Physics 
 Institute, 2~Orlova Rosha, Gatchina, Leningrad district, 188300, Russia}
\newcommand*{\HeJINR}{Veksler and Baldin Laboratory of High Energiy Physics, 
 Joint Institute for Nuclear Physics, 6~Joliot--Curie, Dubna, 141980, Russia}
\newcommand*{\Katow}{August Che{\l}kowski Institute of Physics, University of 
 Silesia, Uniwersytecka~4, 40-007, Katowice, Poland}
\newcommand*{\IFJ}{The Henryk Niewodnicza{\'n}ski Institute of Nuclear 
 Physics, Polish Academy of Sciences, 152~Radzikowskiego St, 31-342 
 Krak\'{o}w, Poland}
\newcommand*{\Tomsk}{Department of Physics, Tomsk State University, 36~Lenina 
 Avenue, Tomsk, 634050, Russia}
\newcommand*{\NuJINR}{Dzhelepov Laboratory of Nuclear Problems, Joint 
 Institute for Nuclear Physics, 6~Joliot--Curie, Dubna, 141980, Russia}
\newcommand*{\KEK}{High Energy Accelerator Research Organisation KEK, Tsukuba, 
 Ibaraki 305--0801, Japan} 
\newcommand*{\ASLodz}{Department of Cosmic Ray Physics, National Centre for 
 Nuclear Research, ul.\ Uniwersytecka~5, 90--950 {\L}\'{o}d\'{z}, Poland}

\author{P.~Adlarson}\altaffiliation[present address: ]{\Mainz}\affiliation{\IKPUU}
\author{W.~Augustyniak} \affiliation{\ASWarsN}
\author{W.~Bardan}      \affiliation{\IPJ}
\author{M.~Bashkanov}   \affiliation{\PITue}\affiliation{\Kepler}
\author{F.S.~Bergmann}  \affiliation{\MS}
\author{M.~Ber{\l}owski}\affiliation{\ASWarsH}
\author{H.~Bhatt}       \affiliation{\IITB}
\author{A.~Bondar}      \affiliation{\Budker}\affiliation{\Novosib}
\author{M.~B\"uscher}\altaffiliation[present address: ]{\PGI,\DUS}\affiliation{\IKPJ}\affiliation{\JCHP}
\author{H.~Cal\'{e}n}   \affiliation{\IKPUU}
% ...

\newcommand*{\Mainz}{Institut f\"ur Kernphysik, Johannes 
 Gutenberg--Universit\"at Mainz, Johann--Joachim--Becher Weg~45, 55128 Mainz, 
 Germany}
\newcommand*{\PGI}{Peter Gr\"unberg Institut, PGI--6 Elektronische 
 Eigenschaften, Forschungszentrum J\"ulich, 52425 J\"ulich, Germany}
\newcommand*{\DUS}{Institut f\"ur Laser-- und Plasmaphysik, Heinrich--Heine 
 Universit\"at D\"usseldorf, Universit\"atsstr.~1, 40225 D??sseldorf, Germany}
\newcommand*{\Bern}{Albert Einstein Center for Fundamental Physics, 
 Universit\"at Bern, Sidlerstrasse~5, 3012 Bern, Switzerland}
\newcommand*{\Aachen}{III.~Physikalisches Institut~B, Physikzentrum, 
 RWTH Aachen, 52056 Aachen, Germany}
\newcommand*{\Delhi}{Department of Physics and Astrophysics, University of 
 Delhi, Delhi--110007, India}
\newcommand*{\Sheff}{Department of Physics and Astronomy, University of 
 Sheffield, Hounsfield Road, Sheffield, S3 7RH, United Kingdom}

\collaboration{WASA-at-COSY Collaboration}\noaffiliation

\begin{abstract}
Exclusive and kinematically complete 
measurements of the double pionic fusion to $^3$He 
have been performed in the energy region of the so-called ABC effect, which
denotes a  pronounced low-mass enhancement in the $\pi\pi$-invariant mass
spectrum. The experiments were carried out with the WASA detector setup
at COSY. Similar to the observations in the basic $pn \to d \pi^0\pi^0$
reaction and in the $dd \to ^4$He$\pi^0\pi^0$ reaction, the data reveal a
correlation between the ABC effect and a resonance-like energy dependence in
the total cross section. Differential cross sections are well described by the
hypothesis of $d^*$ resonance formation during the reaction process in
addition to the conventional $t$-channel $\Delta\Delta$ mechanism. 
The deduced $d^*$ resonance width can be understood from collision broadening
due to Fermi motion of the nucleons in initial and final nuclei.
 
\end{abstract}


\maketitle

\section{Introduction}
\label{intro}

Historically the so-called ABC effect, which denotes an intriguing low-mass
enhancement in the $\pi\pi$ invariant mass spectrum, is known from inclusive
measurements of two-pion  production in nuclear fusion reactions to the
few-body systems $d$, $^3$He and $^4$He. It has been named after the initials of
Abashian, Booth and Crowe, who were the first to observe this effect in 1960 by
studying the inclusive $pd \to ^3$He X reaction \cite{abc}. Its explanation
has been a puzzle since then. In subsequent bubble-chamber \cite{bar,abd} and
single-arm  magnetic spectrometer measurements
\cite{hom,hal,ba,ban,ban1,plo,col,wur,cod} this enhancement was observed also
in double-pionic fusion reactions leading to $d$, $^3$He and $^4$He, if an
isoscalar pion pair was produced. However, such an enhancement was not
observed in fusion reactions leading to deuteron and triton, if an isovector
pion pair was produced. 

These results led to the conclusion that this effect
only appears in reactions, where the participating nucleons fuse to a nuclear
bound system in the final state in combination with the production of an
isoscalar pion pair. 

In recent exclusive and kinematically complete measurements of the $pn \to
d\pi^0\pi^0$ reaction it has been demonstrated \cite{isoabc,prl2011,MB} that
the ABC  
effect in this basic double-pionic fusion reaction is correlated with a narrow
structure in the total cross section with quantum numbers $I(J^P) = 0(3^+)$, a
mass of 2.37 GeV and a width of about 70 MeV. The mass is about 90 MeV below
2$m_{\Delta}$, the mass of a $\Delta\Delta$ system, and the width is three times
narrower than expected from a conventional $t$-channel $\Delta\Delta$ process.

On the contrary the basic isovector fusion process $pp \to d\pi^+\pi^0$ 
exhibits neither an ABC effect nor a narrow resonance structure
\cite{isoabc,FK} in agreement with the observations in all other $pp$ initiated
two-pion channels \cite{hcl,deldel,TT,iso,nnpipi}. 
Isospin decomposition of all three reactions $pn \to d\pi^0\pi^0$, $pn \to
d\pi^+\pi^-$ and $pp \to d\pi^+\pi^0$ leading to the double-pionic
fusion of deuterium ensured that the resonance structure is of purely isoscalar
nature \cite{isoabc}. Also recently published data on the $pn \to
pp\pi^0\pi^-$ reaction show evidence for the resonance structure, though in
this case of an isovector pion pair the ABC effect is absent
\cite{pppi0pi-}. Compelling evidence that the isoscalar resonance structure
observed in two-pion production processes denotes truly a $s$-channel
resonance in the $pn$ system comes from polarized $np$ scattering in the
energy region of the ABC effect \cite{nppol,nppolfull}. Inclusion of these
data in the SAID data base with subsequent partial-wave analysis produces a
pole at ($2380\pm10 - i40\pm5$) MeV in the coupled $^3D_3 - ^3G_3$ partial
waves in full agreement with the here discussed resonance hypothesis. 

Since in these latter
reactions the resonance is not associated with any ABC effect, it was called no
longer ABC resonance, but $d^*$ \cite{pppi0pi-} --- in historical reference to
a predicted \cite{goldman,ping} dibaryon with exactly the quantum numbers as
we observe it now.  

The existence of the ABC effect in the double-pionic fusion to $^3$He and
$^4$He  has been confirmed by exclusive and kinematically complete
experiments at CELSIUS/WASA \cite{mb,SK} and recently also at ANKE-COSY
\cite{anke}. In measurements at WASA-at-COSY it
has been additionally shown that in the $dd \to ^4$He$\pi^0\pi^0$ reaction the
ABC effect is again correlated with a resonance structure in the total cross
section at $\sqrt s \approx$~2.37 GeV + 2$m_N$  \cite{AP}. However, in
comparison to the  basic fusion reaction to deuterium 
the width of the resonance structure appears substantially broadened, which
may be attributed to the Fermi motion of the nucleons in initial and final
nuclei as well as due to collision damping.  

So what is left in this scenario is the question, whether also in case of the
double-pionic fusion to $^3$He the ABC effect is correlated with a resonance
structure in the total cross section.

\section{Experiment}
\label{sec:1}

In an effort to find an experimental answer for this question we have analyzed
corresponding two-pion production data, which were  obtained with WASA at COSY 
\cite{barg,wasa} primarily for other reasons. The data sets, which we used,
originate from two different runs. 

The first run concerns a proton beam of energy $T_p$ = 1.0 GeV
hitting the deuterium pellet target \cite{barg,wasa}. This allows us to
analyze the reaction $pd \to ^3$He$\pi^0\pi^0$ at $T_p$ = 1.0 GeV. The beam
energy corresponds to a center-of-mass (cm) energy of $\sqrt s$ = 3.42 GeV =
2.48 GeV + $m_N$, {\it i.e.} pertains to the high-energy end of
the region, where the ABC effect has been observed previously
\cite{abc,bar,abd,mb}.

The second data set used for our purposes concerns runs with a deuteron beam
of $T_d$ = 1.4 and 1.7 GeV, respectively, hitting the deuterium pellet
target. We use these runs to obtain data for the quasifree reaction $dd \to
^3$He$\pi^0\pi^0 + n_{spectator}$ in the range 3.1 GeV $< \sqrt s <$ 3.4 GeV
(with respect to the $^3$He$\pi^0\pi^0$ system), {\it i.e.} covering just the
ABC region. 

Both data sets allow an exclusive and kinematically complete reconstruction of
the $^3$He$\pi^0\pi^0$ events with kinematic overconstraints.

The trigger for a valid event was just a single track in the forward detector
of WASA  with high thresholds in its first scintillation detector layers, in
order to suppress fast protons and deuterons. With this trigger condition the
data rate of accepted events was at moderate 2
kHz. The selection criteria for the offline analysis were a single He track in
the forward detector and four neutral hits in the central detector.

The emerging $^3$He particles were 
registered in the forward detector of WASA and 
identified by the $\Delta$E-E technique. The photons from the $\pi^0$
decay were detected and identified in the central detector \cite{barg}. 
Consequently four-momenta were measured for all emitted particles of an
event with the exception of the spectator neutron, which appears in the second
reaction type only. 

Together with the condition that two pairs of the detected photons have 
to fulfill the $\pi^0$ mass condition, we have 6 overconstraints for the
kinematic fit of an event in the first reaction type and 3 overconstraints in
the second case. From the three possible combinations to reconstruct
the four-momenta of the two pions out of four photon signals the one with the
smallest $\chi^2$ has been selected \cite{AP,TT}.

All particles have been detected over the
full solid angle with the exception of those $^3$He ejectiles, which escaped
in the beam-pipe (polar angles $\Theta_{^3He}^{lab} < 3^\circ$). 

Acceptance and efficiency corrections have been made by use of Monte Carlo (MC)
simulations of detector setup and performance. For a self-consistent procedure
the reaction model used in the MC simulations has been iteratively adjusted to
the experimental results.

With regard to the second data set the momentum spectra of the reconstructed
neutron are shown in Fig.~1, at the top for  $T_d$ = 1.4 GeV and at the bottom
for  $T_d$ = 1.7 GeV. The strong 
enhancement of events at low momenta corresponds to the situation, when the
spectator neutron originates from the target deuteron, whereas the enhancement
at the high-momentum end corresponds to a spectator neutron stemming from a beam
deuteron. The area in between is covered by non-quasifree processes, so-called
coherent processes, where the reconstructed neutron is not just a kinematic
spectator, but also plays an active role in the reaction
dynamics. Misidentified $^4$He particles could be eliminated in subsequent
analysis steps by the constraint
that the reconstructed neutron should not have the same direction and velocity
as the detected He particle.

\begin{figure} 
\centering
\includegraphics[width=0.9\columnwidth]{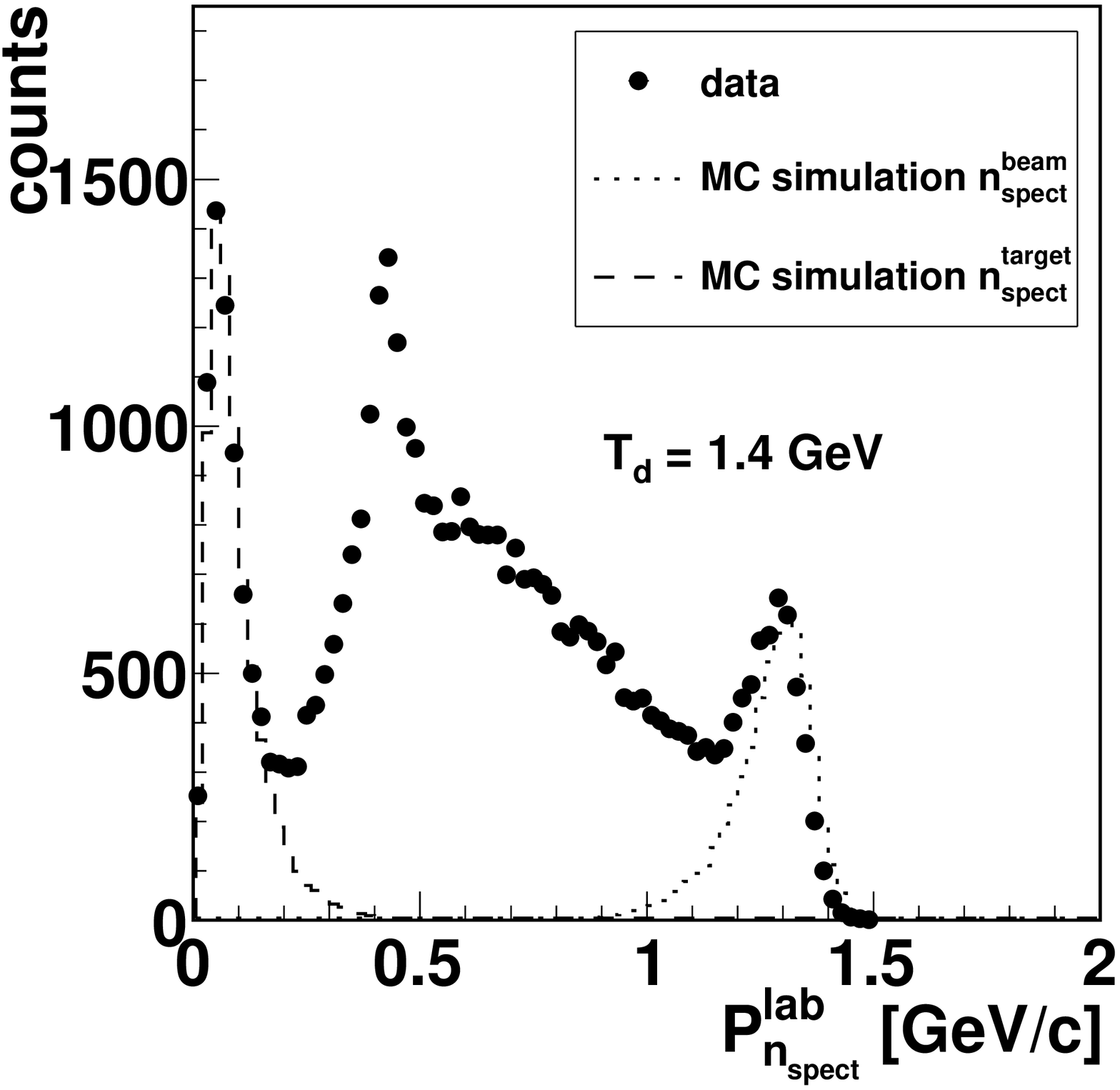}
\includegraphics[width=0.9\columnwidth]{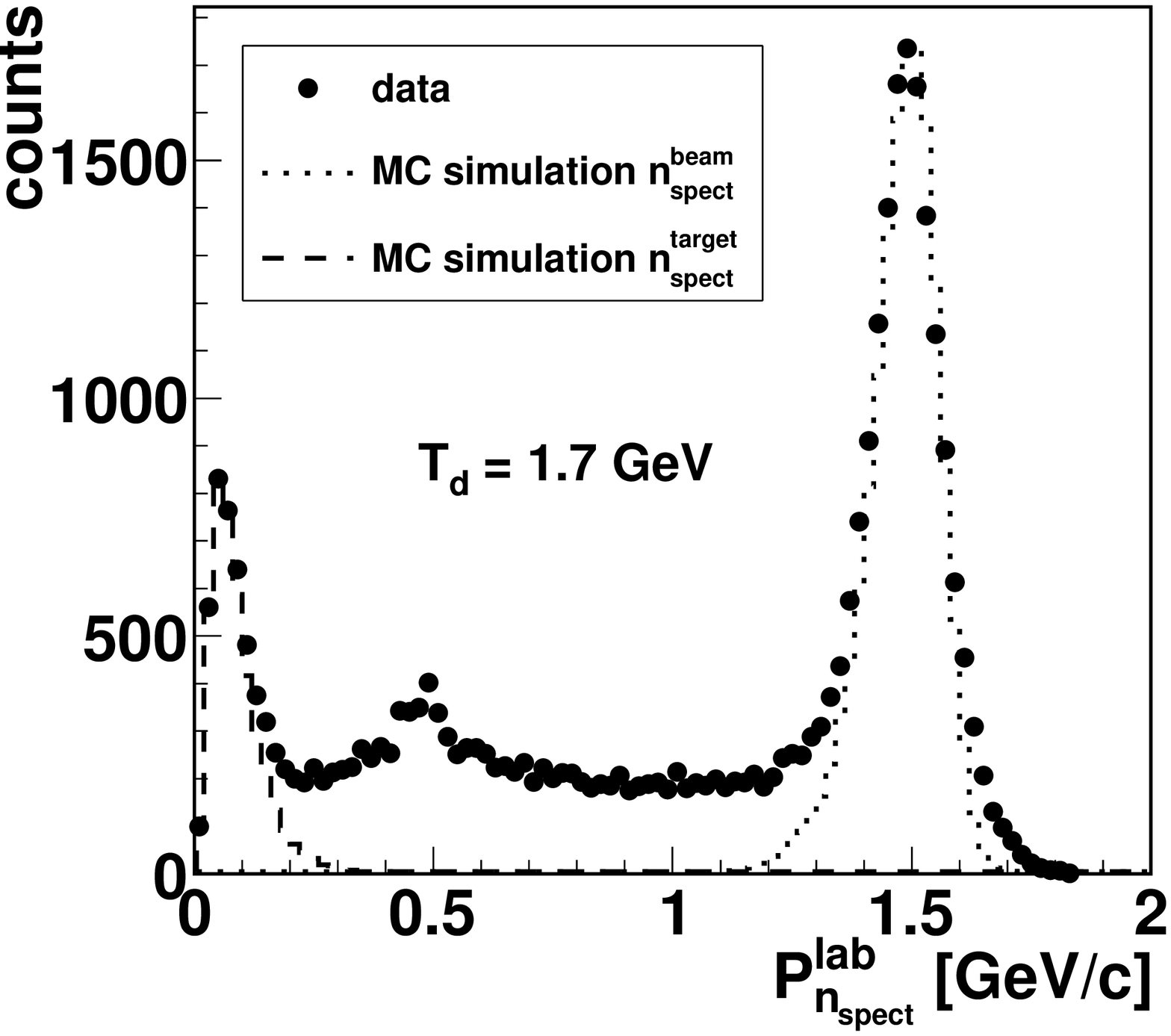}
\caption{\small 
  Distribution of the reconstructed neutron momenta in the $dd \to
  ^3$He$\pi^0\pi^0 + n$ reaction at $T_d$ = 1.4 GeV (top) and 1.7 GeV
  (bottom), respectively. Data are given by solid dots. The dashed
  (dotted) lines show 
  the expected distribution for the quasifree process based on the CD Bonn 
  potential \cite{mach} deuteron wavefunction, if the spectator originates
  from the target (beam) deuteron. 
  The peak near 0.5 GeV/c originates from $^4$He contamination, which has been
  removed in subsequent analysis steps. 
}
\label{fig1}
\end{figure}

\begin{figure} 
\centering
\includegraphics[width=.9\columnwidth]{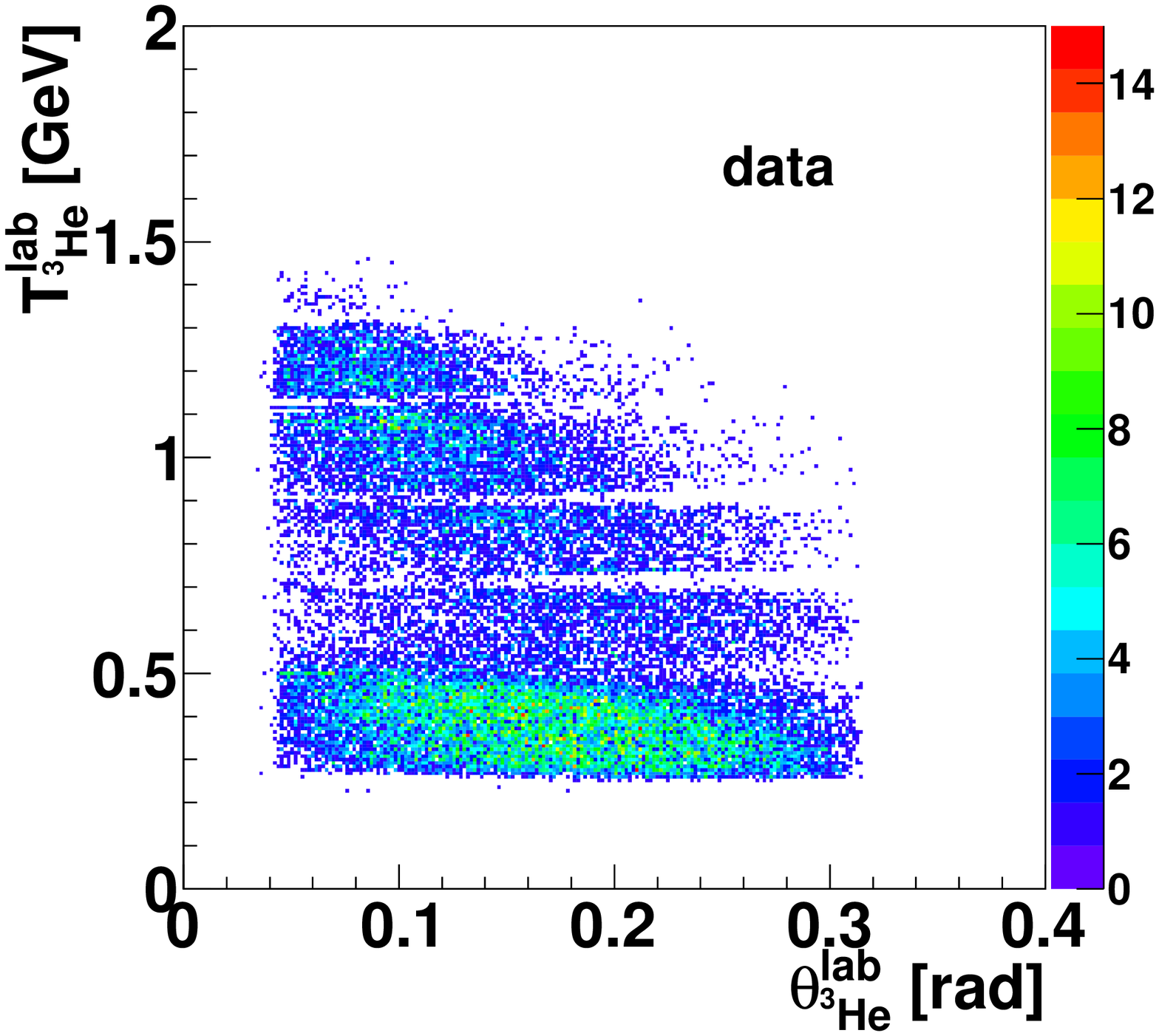}
\includegraphics[width=.9\columnwidth]{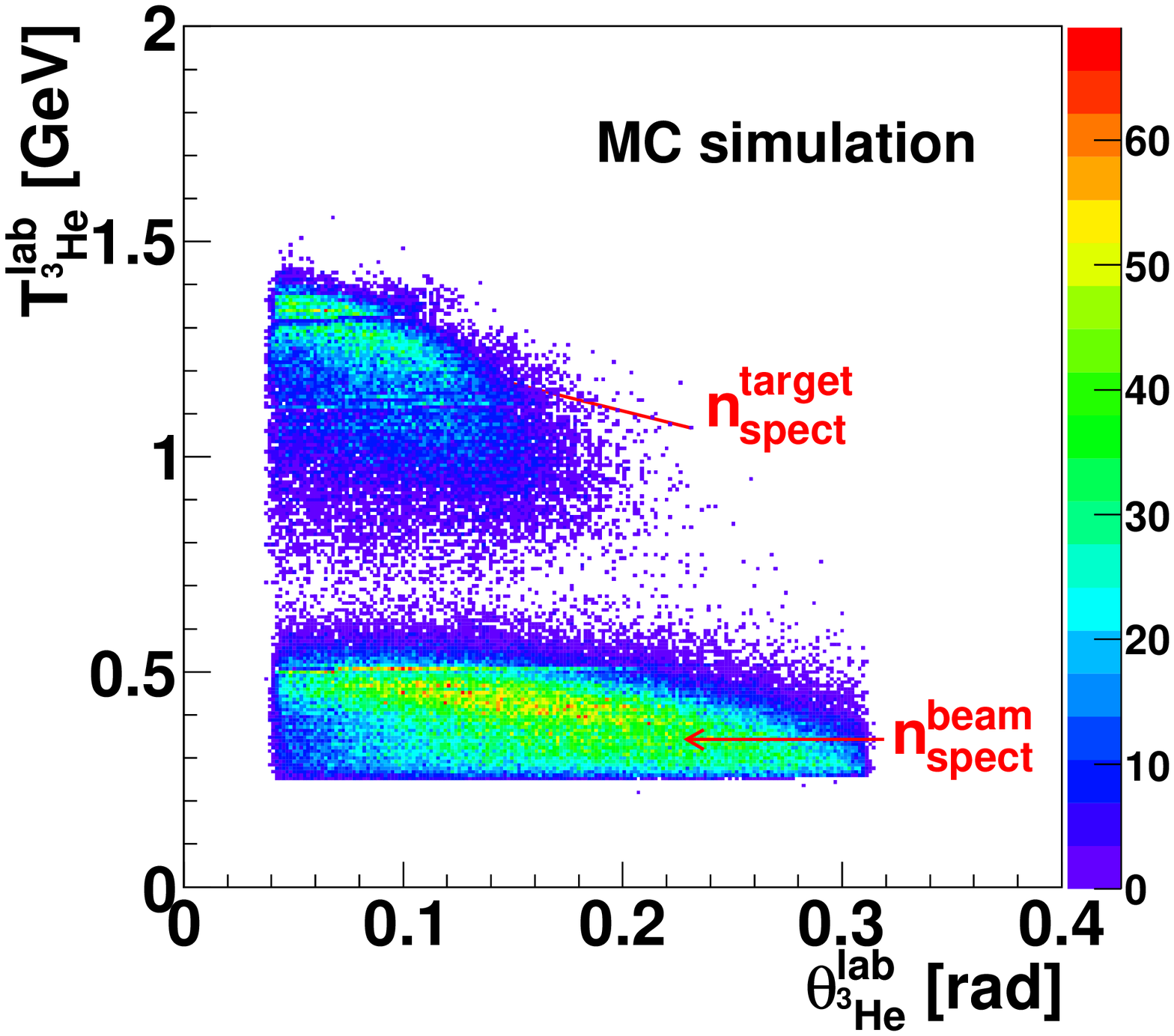}
\caption{\small 
  Scatter plot of the kinetic energy of the $^3$He ejectiles in the $dd \to
  ^3$He$\pi^0\pi^0 + n$ reaction at $T_d$ = 1.7 GeV versus the $^3$He lab
  scattering angle $\Theta_{^3He}^{lab}$. At the top data are shown and at the
  bottom the 
  corresponding MC simulation of the processes with either the neutron
  spectator in the target or in the beam. The high-density area at small
  angles and large kinetic energies 
  corresponds to the process with the spectator neutron in the
  target, whereas the distribution at small energies and large 
  scattering angles belongs to spectator neutrons in the beam. The area
  in between corresponds to coherent processes. 
}
\label{fig2}
\end{figure}

In case of a target spectator neutron ($dp$ reaction) the emitted $^3$He
particles are at very small forward angles due to the Lorentz boost, so that
the lower limit of $\Theta_{lab} \geq 3^\circ$ cuts
severely into the reaction phase space rendering acceptance corrections largely
model-dependent. Hence we refrain from giving results for this scenario. 

For
the case of the neutron spectator originating from the beam deuteron ($pd$
reaction) the
situation is kinematically much more favorable. Unfortunately also here we
met difficulties, since the energies of the $^3$He ejectiles deposited in the
segments of the forward detector turned out to be partly below the trigger
thresholds, which were increased for the observation of other reaction
channels of primary interest in these runs. As a consequence we had to
tune the actual trigger thresholds individually for each of the detector
segments very carefully by adjusting the simulations of the detector
performance to the observed response of each of the corresponding
scintillation detectors. 

Comparing the spectra in Fig.~1, top and bottom, we see that at higher beam
energy the coherent process has much reduced compared to the quasifree
process. Hence, also in this respect it is more favorable to analyze the 1.7
GeV data for the $^3$He$\pi^0\pi^0$ production channel.

Fig.~2 displays the two-dimensional scatter plot of the kinetic energy of the
$^3$He ejectile versus its polar scattering angle in the lab
system. 
The $^3$He particles originating from the
quasifree process in the target deuteron ($dp$ reaction) produce a strong
enhancement at small angles in combination with large kinetic energies in the
scatter plot, whereas 
the $^3$He ejectiles from the quasifree process in the beam deuteron ($pd$
reaction) produce a
strong enhancement at small energies over a wide region of angles. In the
scatterplot these two regions are strongly populated and well separated from
the region in between, which covers coherent processes. In order to get rid of
the latter as well as of the target related spectators, we subsequently
constrain the polar angle for the reconstructed neutrons to the kinematical
spectator limit of $\Theta_n^{cm} \leq 11.5^\circ$ for beam related spectators,
where the superscript $cm$ denotes the angle in the $dp$ center-of-mass
system. That way we obtain a momentum spectrum of the spectator neutrons,
which is very close to that given by the
dotted line in Fig.~1 and which is essentially free of background.

The absolute normalization of the data from the single-energy measurement at
$T_p$ = 1.0 GeV was obtained by a
relative normalization to the $pd \to ^3$He$\pi^0$ reaction
measured simultaneously with the same trigger. Our results for this reaction
in turn have been normalized to those from Saclay measurements at neighboring
energies \cite{kerboul,banaigs}. 
Though this procedure appears to be straightforward, it contains a number of
difficulties. The Saclay data appear to be most reliable at $\Theta_{^3He} =
180^\circ$ \cite{kerboul}, where WASA can not measure. 
Hence we used the full back-angle hemisphere to adjust the WASA results to
those of Saclay. However, due to the scarcity of Saclay data at finite angles
we estimate that the total  
uncertainty in the absolute normalization could be as large as 30$\%$. For
details see Ref. \cite{EP}. 

The data of the quasi-free run overlap with the single-energy measurement
at their high-energy end. Hence, for simplicity they have been normalized to
the result of the single-energy measurement.

\section{Results}

Resulting observables of the normalized as well as acceptance and
efficiency corrected data are displayed in Figs.~3 - 6. 

The total cross section data obtained from the analysis of both experiments are
shown in Fig.~3, which exhibits the energy dependence of the total cross
section for the $^3$He$\pi^0\pi^0$ production. Our result from the
run at $T_p$ = 1.0 GeV ($\sqrt s$ = 3.416 GeV) is shown by the filled triangle
symbol, whereas the results from the quasifree run are given by the filled
circles. The shaded area denotes the estimated systematic uncertainties, which
result dominantly from the efficiency and acceptance corrections. Also
uncertainties from rest gas contributions and kinematic fit are contained in
this estimate.

Included in Fig.~3 are also the results from previous exclusive
measurements at CELSIUS-WASA at $T_p$ = 0.893 GeV (open square) \cite{mb} and
at PROMICE/WASA at $T_p$ = 0.477 GeV (open circle) \cite{Anders}, the latter
carried out at CELSIUS, too. 
In order to
avoid systematic discrepancies in the procedure used for the absolute
normalization, the CELSIUS-WASA result has been reanalyzed by subjecting it to
exactly the same procedure ({\it i.e.}, considering the full back-angle
hemisphere) as applied now for the single-energy measurement at
$T_p$ = 1.0 GeV. As 
a result the CELSIUS-WASA value at $T_p$ = 0.893 GeV changed from the
published value of 2.8(3) $\mu$b \cite{MB} to 1.9(3) $\mu$b with the latter
value being plotted
in Fig.~3. The revised value is in good agreement with the new data. We note
that the COSY-ANKE result for the $pd \to ^3$He$\pi^+\pi^-$ reaction is also
lower by 40$\%$ \cite{anke}, when compared to the corresponding published
value from CELSIUS-WASA -- in agreement with our finding for the
$^3$He$\pi^0\pi^0$ channel.

\begin{figure} 
\includegraphics[width=0.99\columnwidth]{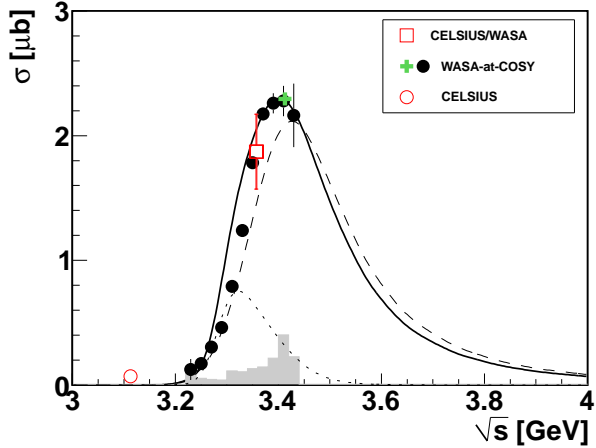}
\caption{Energy dependence of the total cross section for the
  double-pionic fusion to $^3$He with the production of a $\pi^0\pi^0$
  pair. Data obtained in this work by measurements of the $pd \to
  ^3$He$\pi^0\pi^0$ reaction at $T_p$ = 1.0 GeV and of the $dd \to
  ^3$He$\pi^0\pi^0 + n_{spectator}$ reaction at $T_d$ = 1.7 GeV are given by
  the filled cross and the filled circles, respectively. They are
  compared to previous results from PROMICE/WASA \cite{Anders} 
  (open circle) 
   and  CELSIUS/WASA \cite{mb} (open square). The latter has been
   renormalized, see text.
  The shaded area denotes the estimated systematic uncertainties. The dotted
  curve gives the $d^*$ contribution, the dashed line the $t$-channel $\Delta
  \Delta$ process and the solid line their (coherent) sum.
}
\label{fig:3} 
\end{figure}

\begin{figure} 
\centering
\includegraphics[width=0.7\columnwidth]{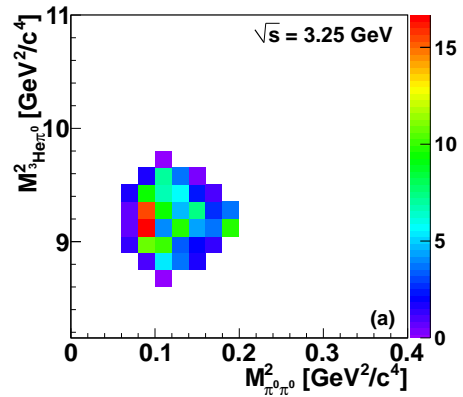}
\includegraphics[width=0.7\columnwidth]{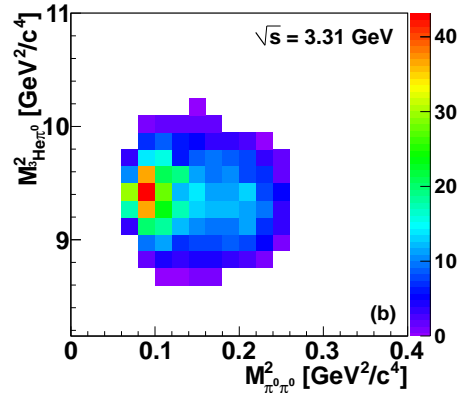}
\includegraphics[width=0.7\columnwidth]{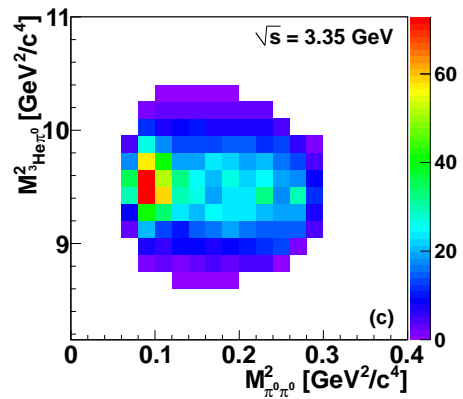}
\includegraphics[width=0.7\columnwidth]{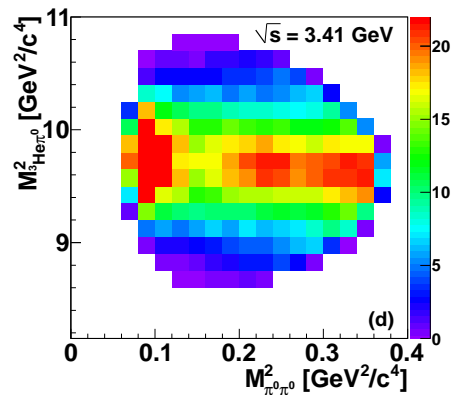}

\caption{\small
  Dalitz plots of $M_{^3He\pi^0}^2$ versus $M_{\pi^0\pi^0}^2$ of the data at cm
  energies of $\sqrt{s}$ = 3.25 GeV, $\sqrt{s}$ = 3.31 GeV, $\sqrt{s}$ = 3.35
  GeV and $\sqrt{s}$ = 3.41 GeV (from top to bottom).  
}
\label{fig:4}
\end{figure}

\begin{figure} [t]
\centering
\includegraphics[width=0.49\columnwidth]{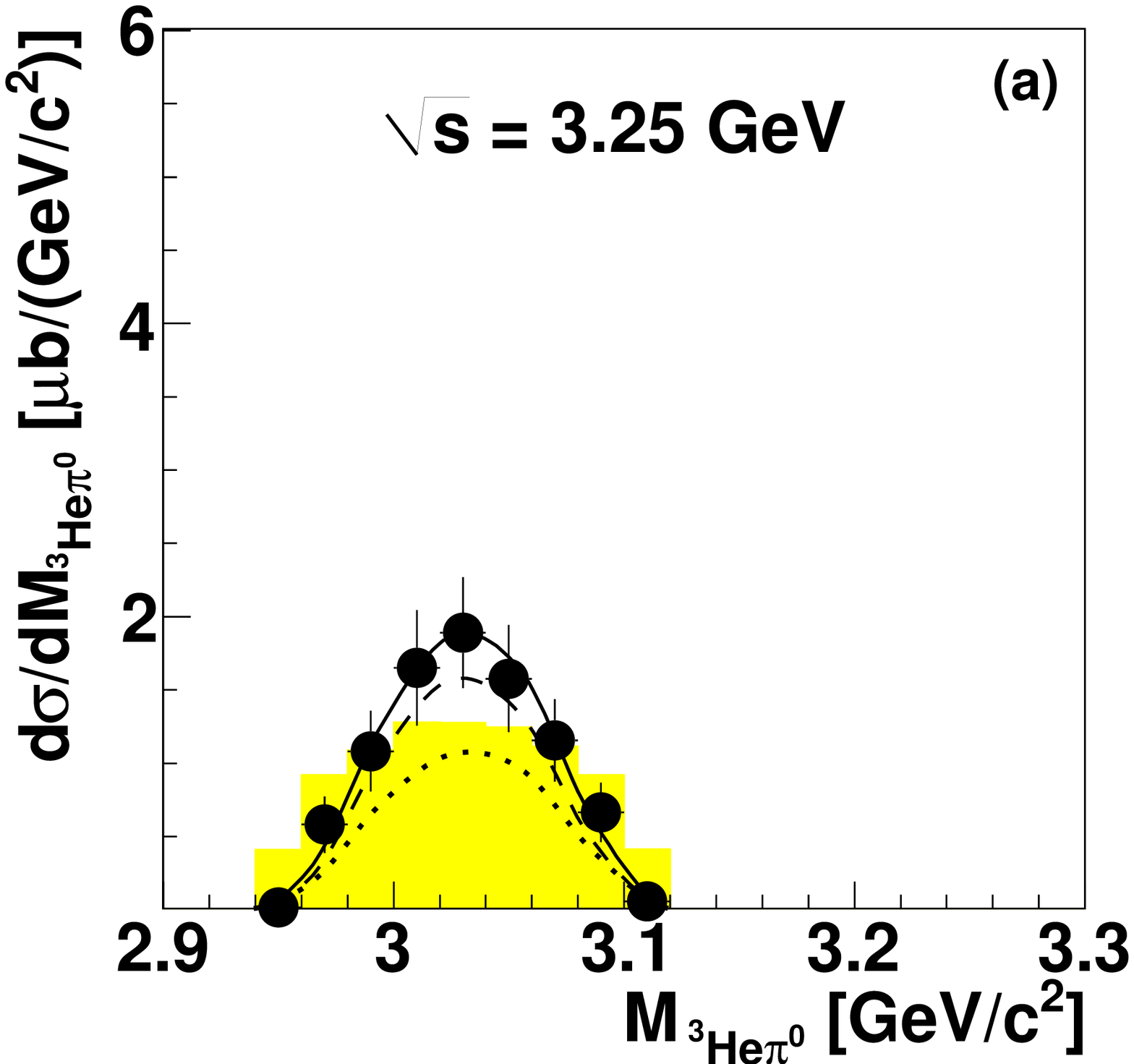}
\includegraphics[width=0.49\columnwidth]{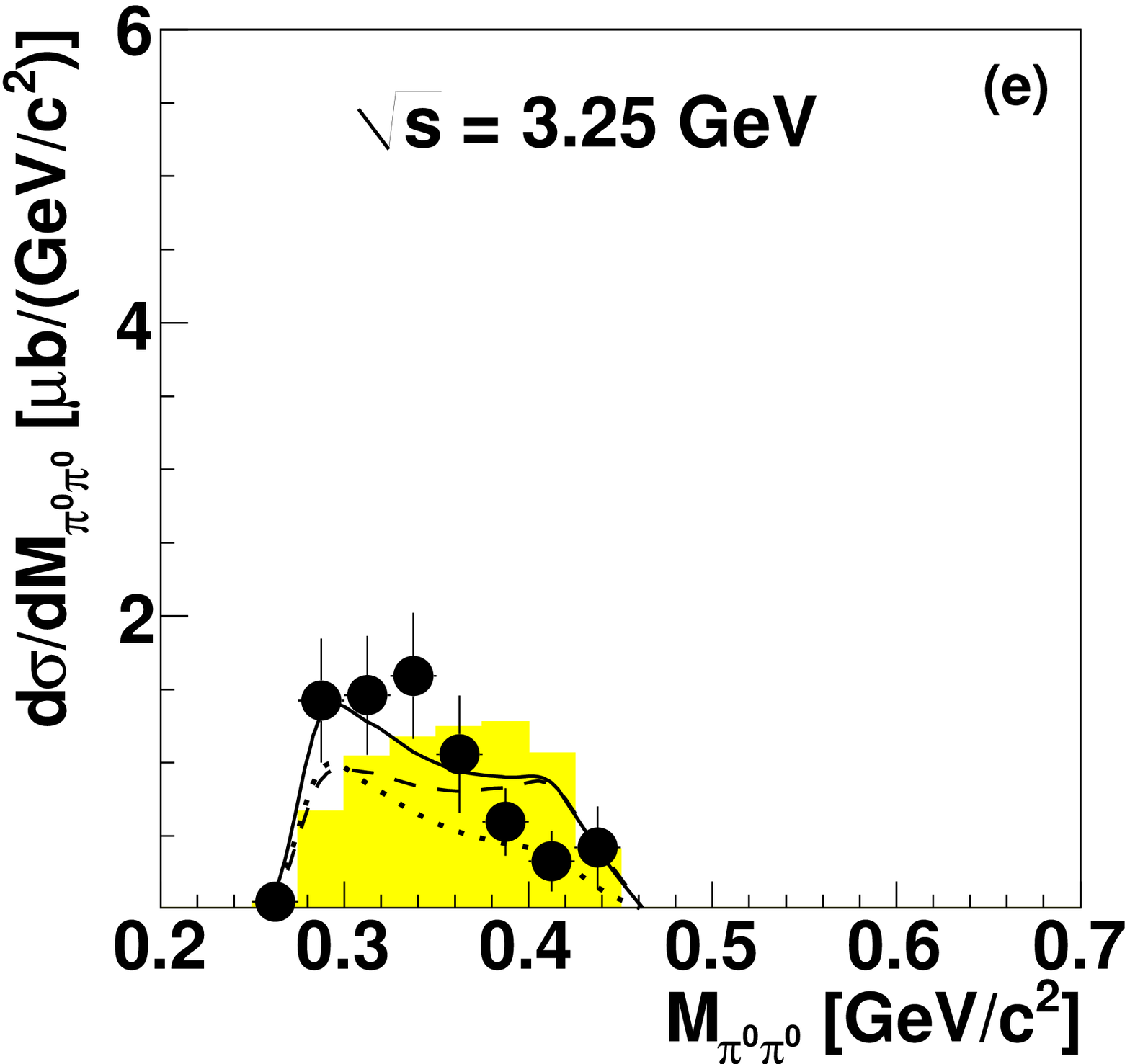}
\includegraphics[width=0.49\columnwidth]{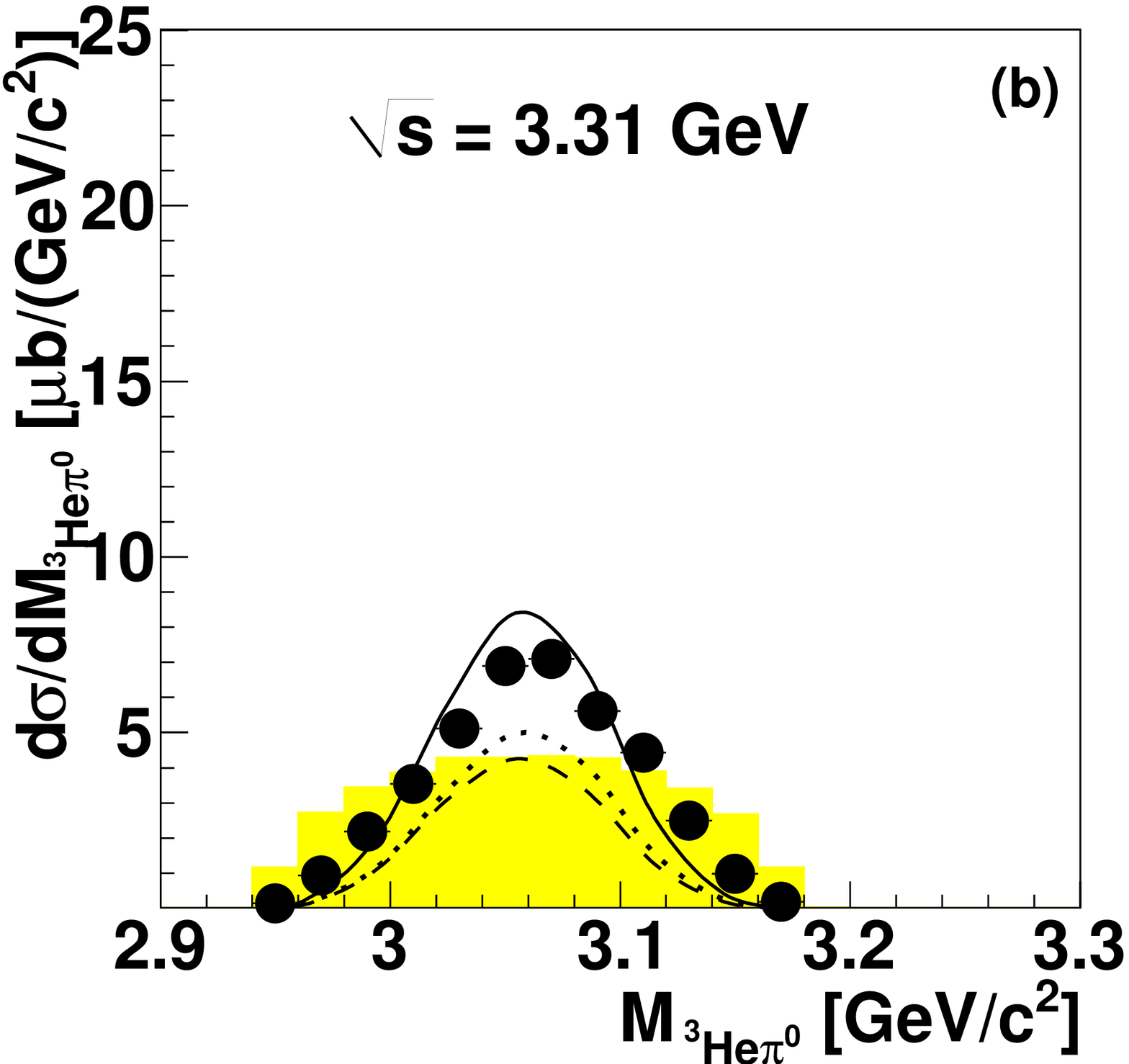}
\includegraphics[width=0.49\columnwidth]{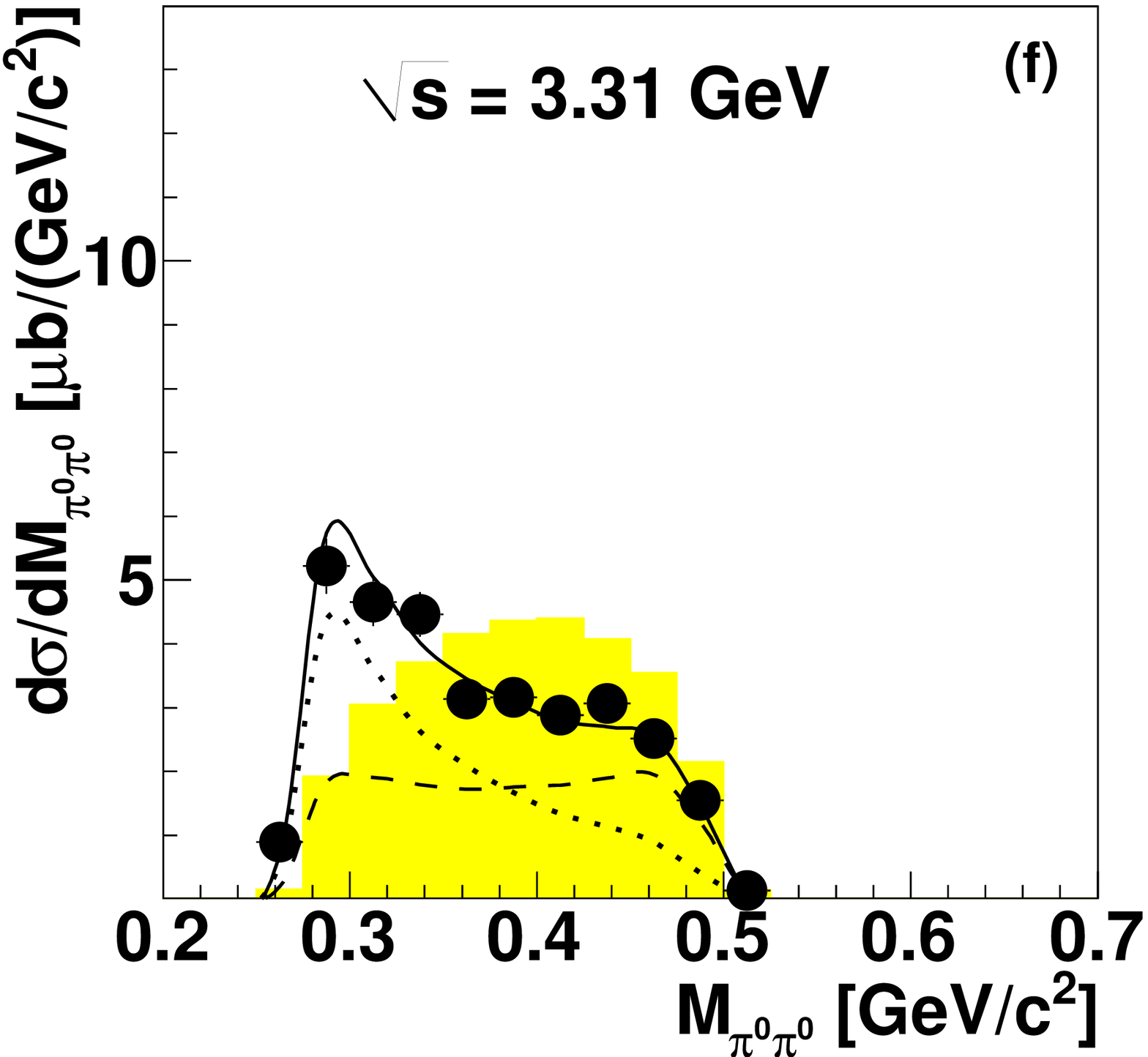}
\includegraphics[width=0.49\columnwidth]{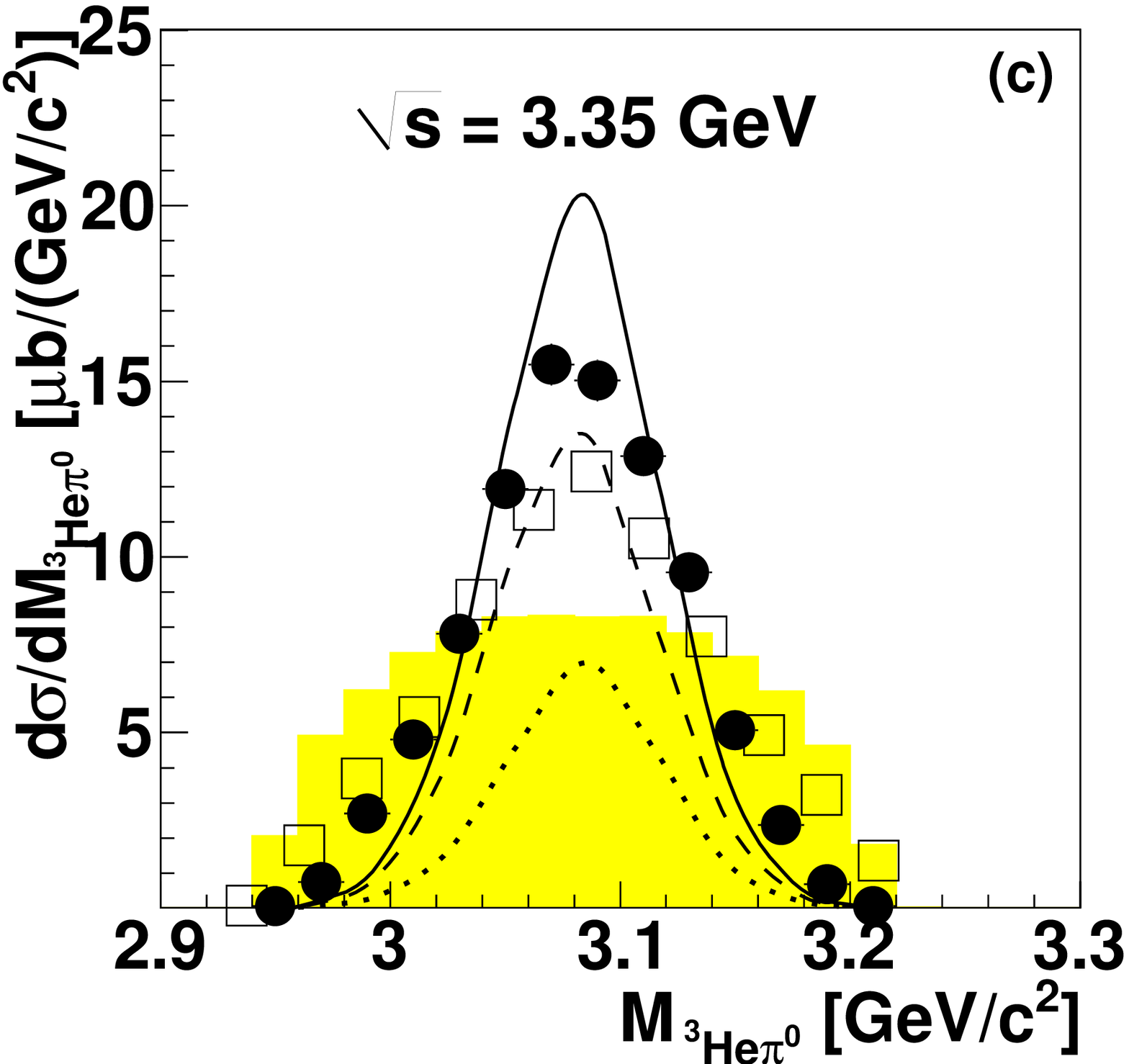}
\includegraphics[width=0.49\columnwidth]{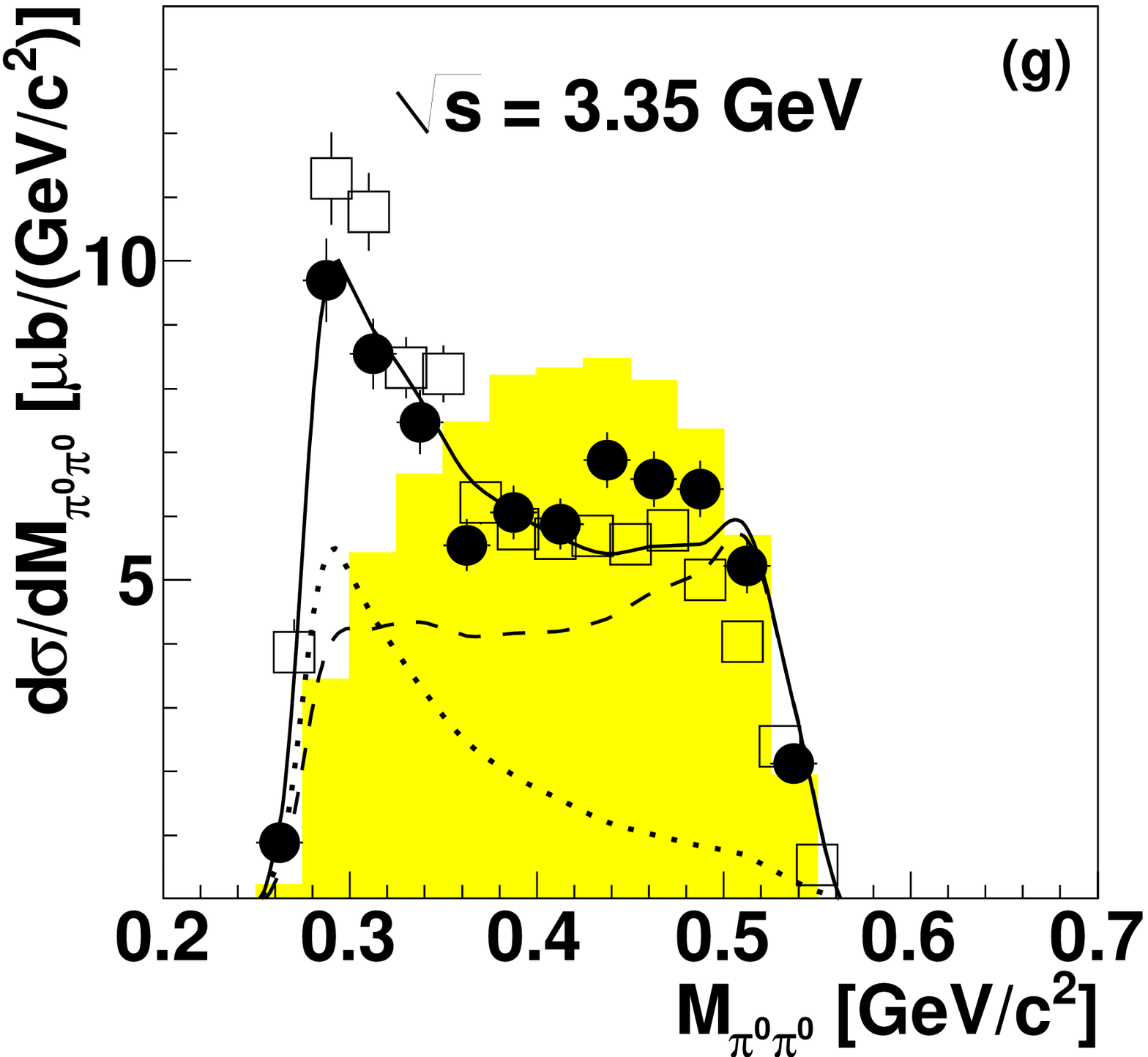}
\includegraphics[width=0.49\columnwidth]{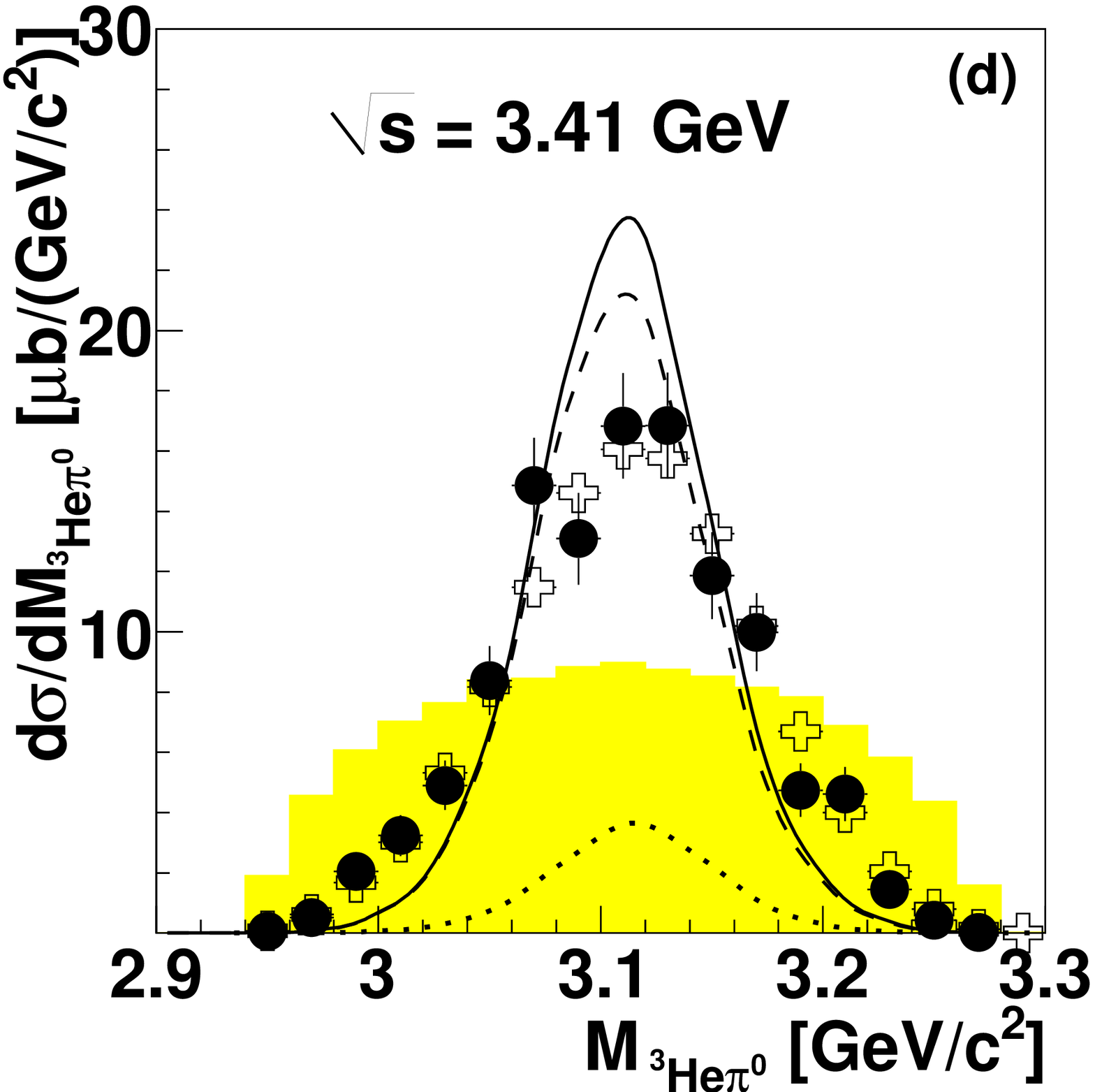}
\includegraphics[width=0.49\columnwidth]{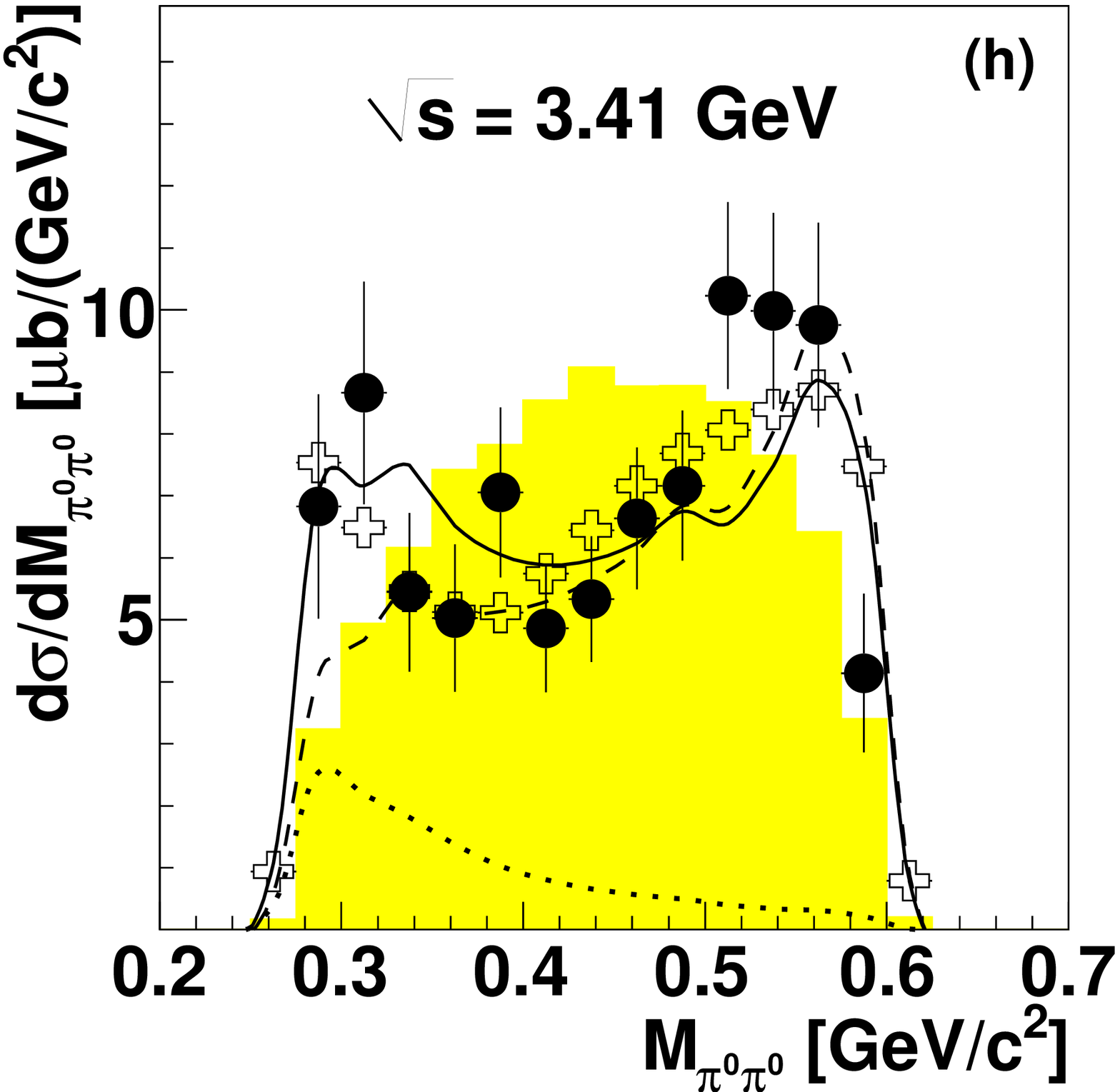}
\caption{\small 
  Distributions of $M_{^3He\pi^0}$ ({\bf left}) and $M_{\pi^0\pi^0}$ ({\bf right})
  at $\sqrt{s}$ = 3.25, 3.31, 3.35 and 3.41 GeV (from {\bf top} to {\bf
    bottom}). Filled circles denote data from the quasifree runs, open crosses
  those from the $pd$ reaction at $T_p$ = 1.0 GeV ($\sqrt s$ = 3.416 GeV). 
  Data from CELSIUS/WASA at $T_p$ = 0.89 
  GeV ($\sqrt s$ = 3.35 GeV) \cite{mb} are shown by open squares. The shaded
  area denotes the phase-space distribution. The dotted curve gives the $d^*$
  contribution, the dashed line the $t$-channel $\Delta\Delta$ process and the
  solid line their sum.
}
\label{fig:5}
\end{figure}

\begin{figure} 
\begin{center}
\includegraphics[width=0.49\columnwidth]{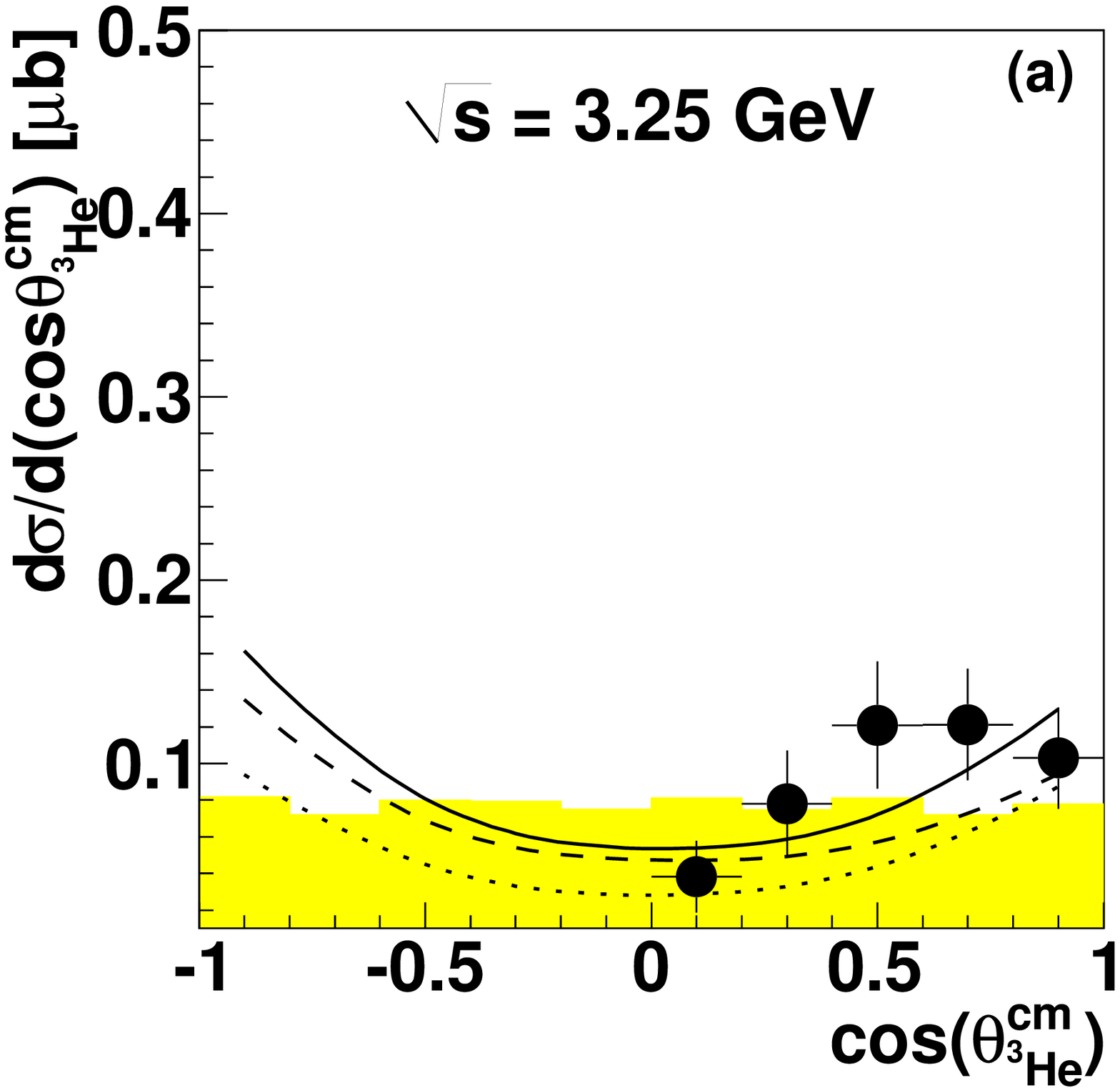}
\includegraphics[width=0.49\columnwidth]{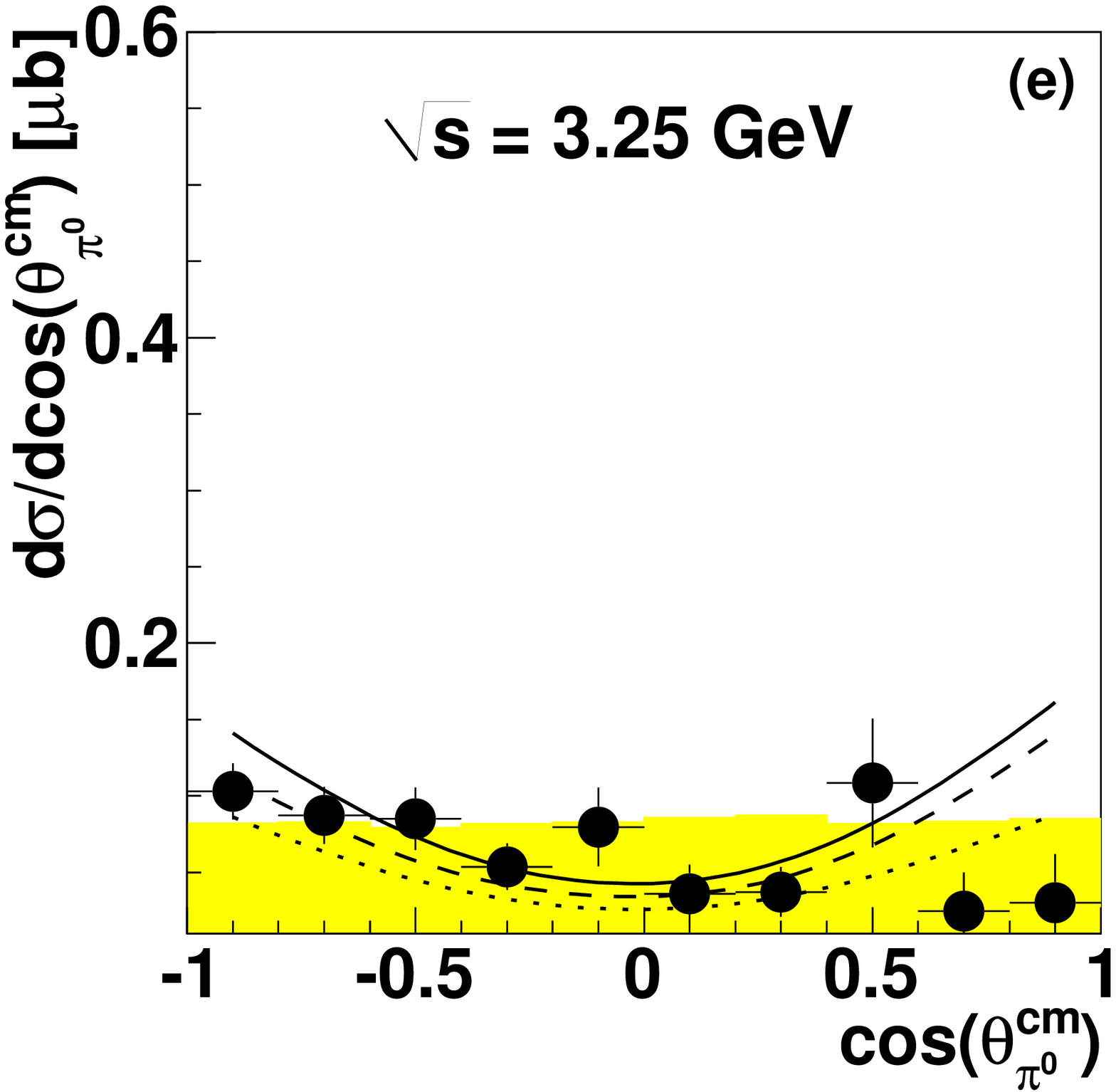}
\includegraphics[width=0.49\columnwidth]{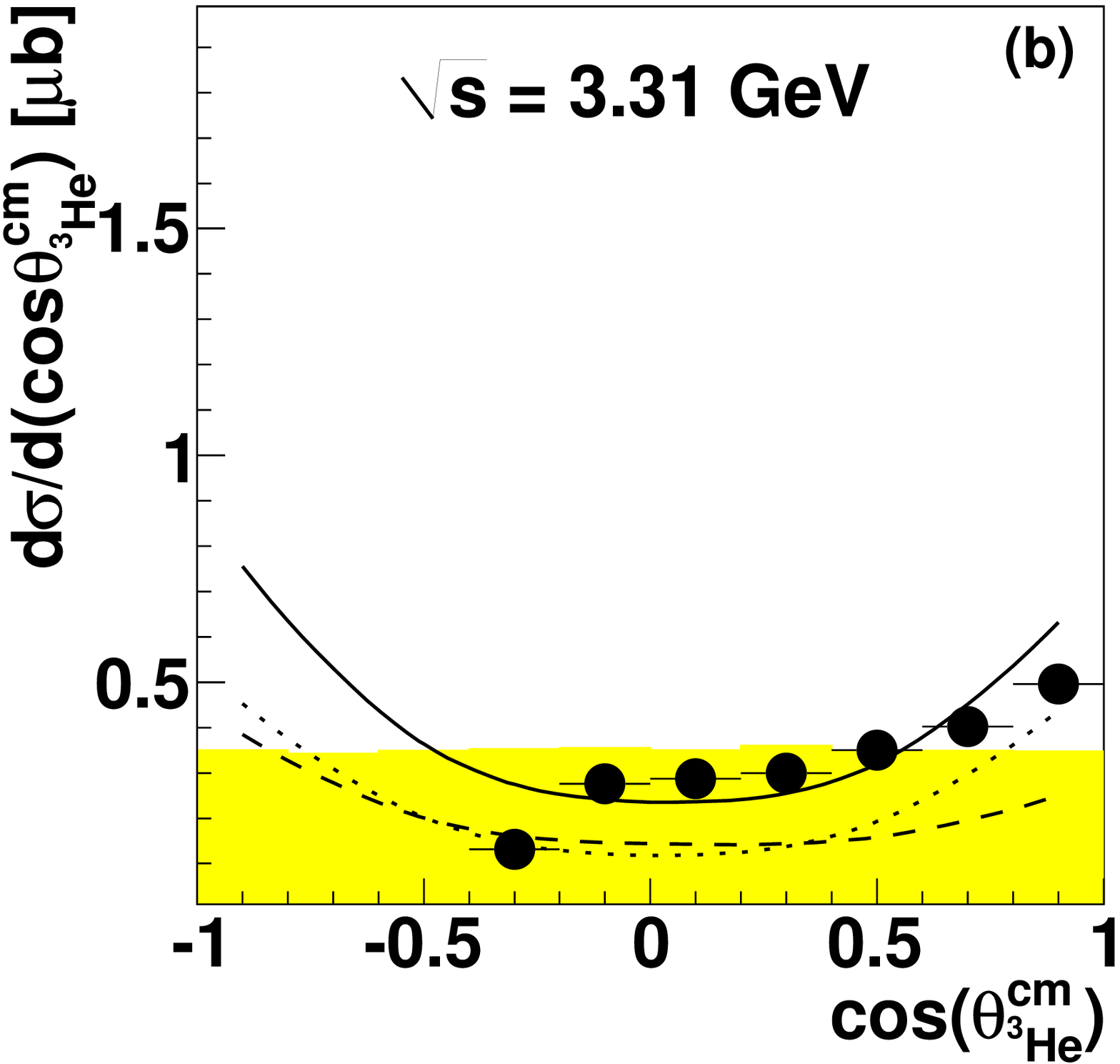}
\includegraphics[width=0.49\columnwidth]{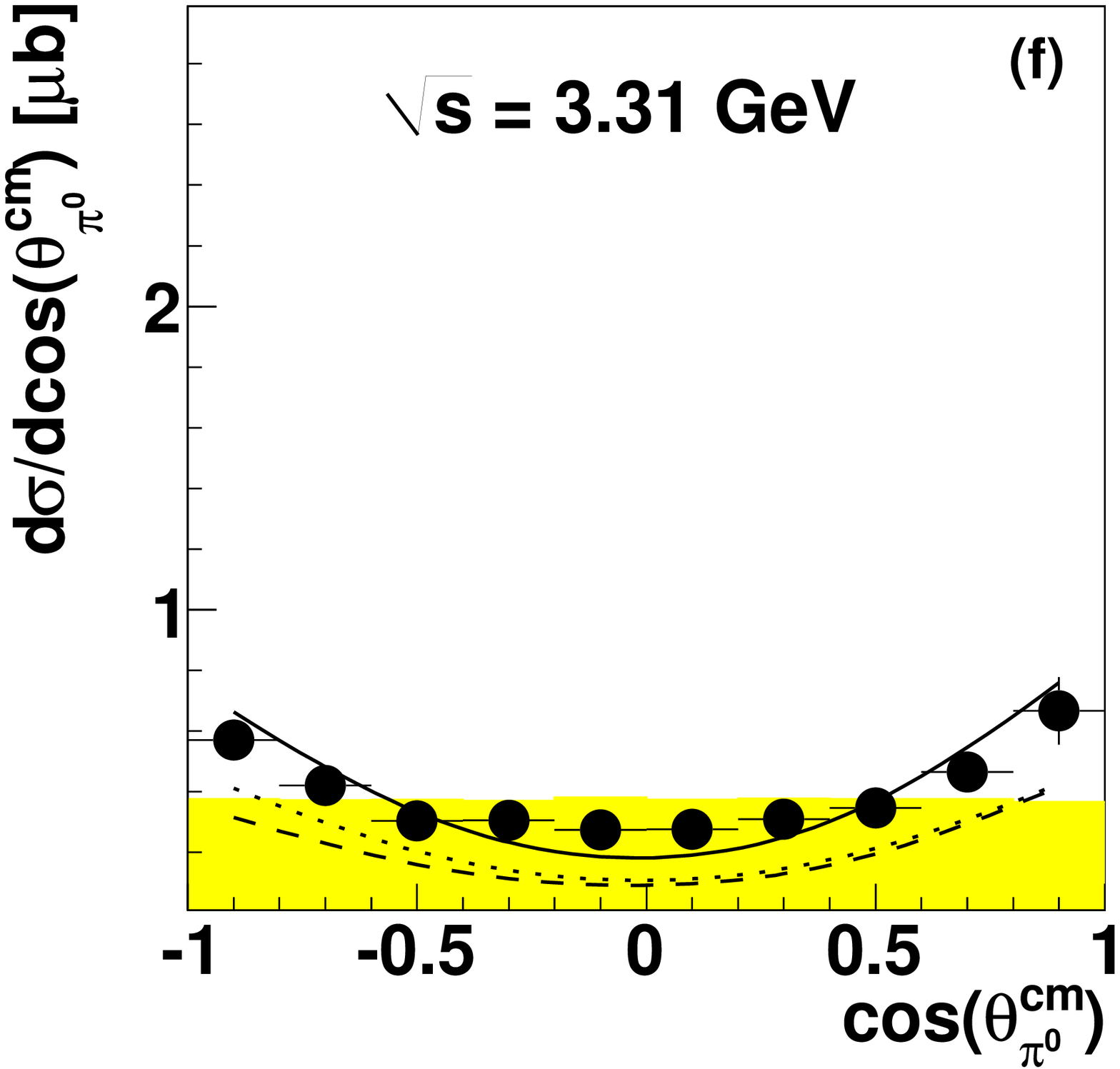}
\includegraphics[width=0.49\columnwidth]{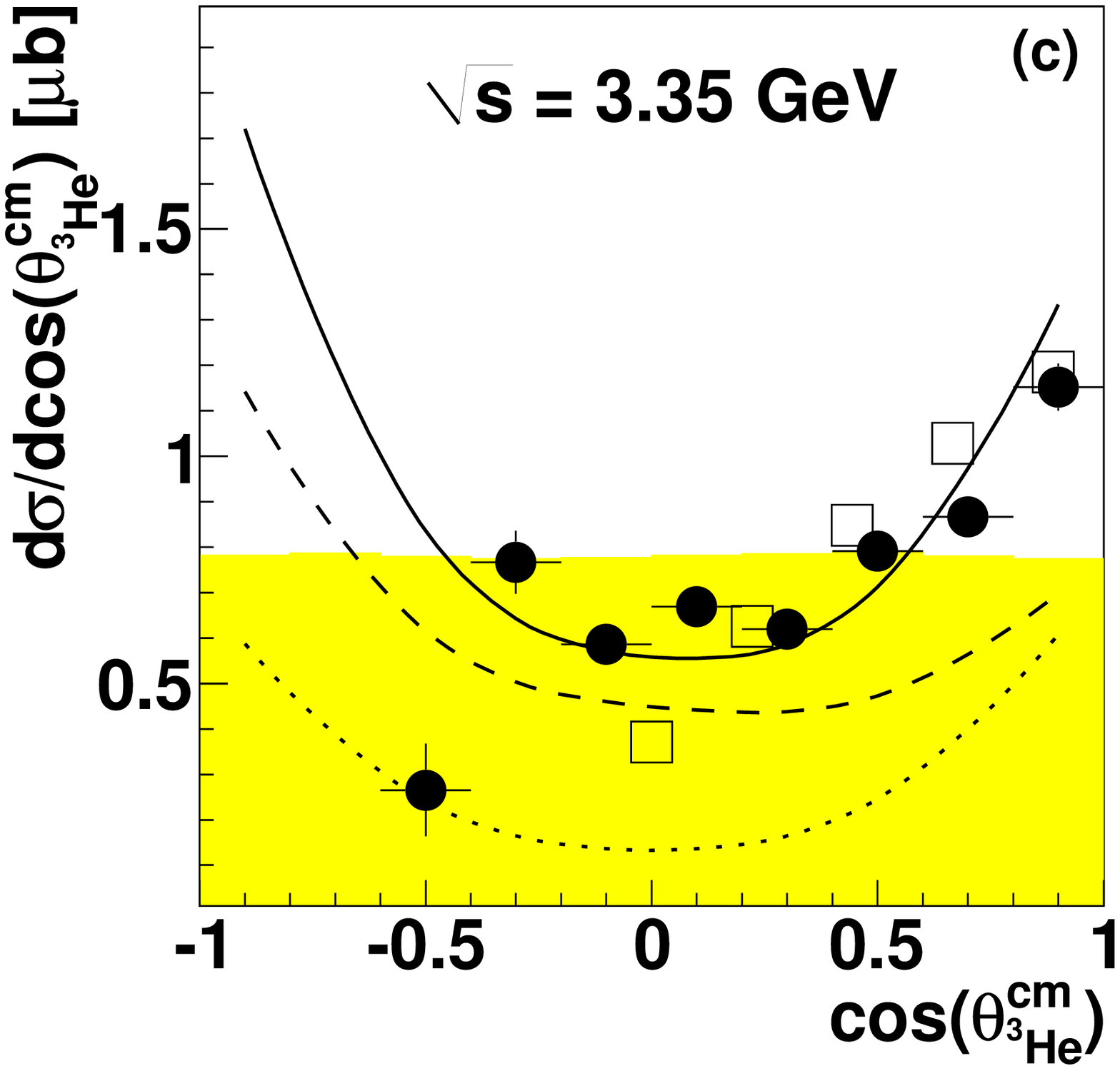}
\includegraphics[width=0.49\columnwidth]{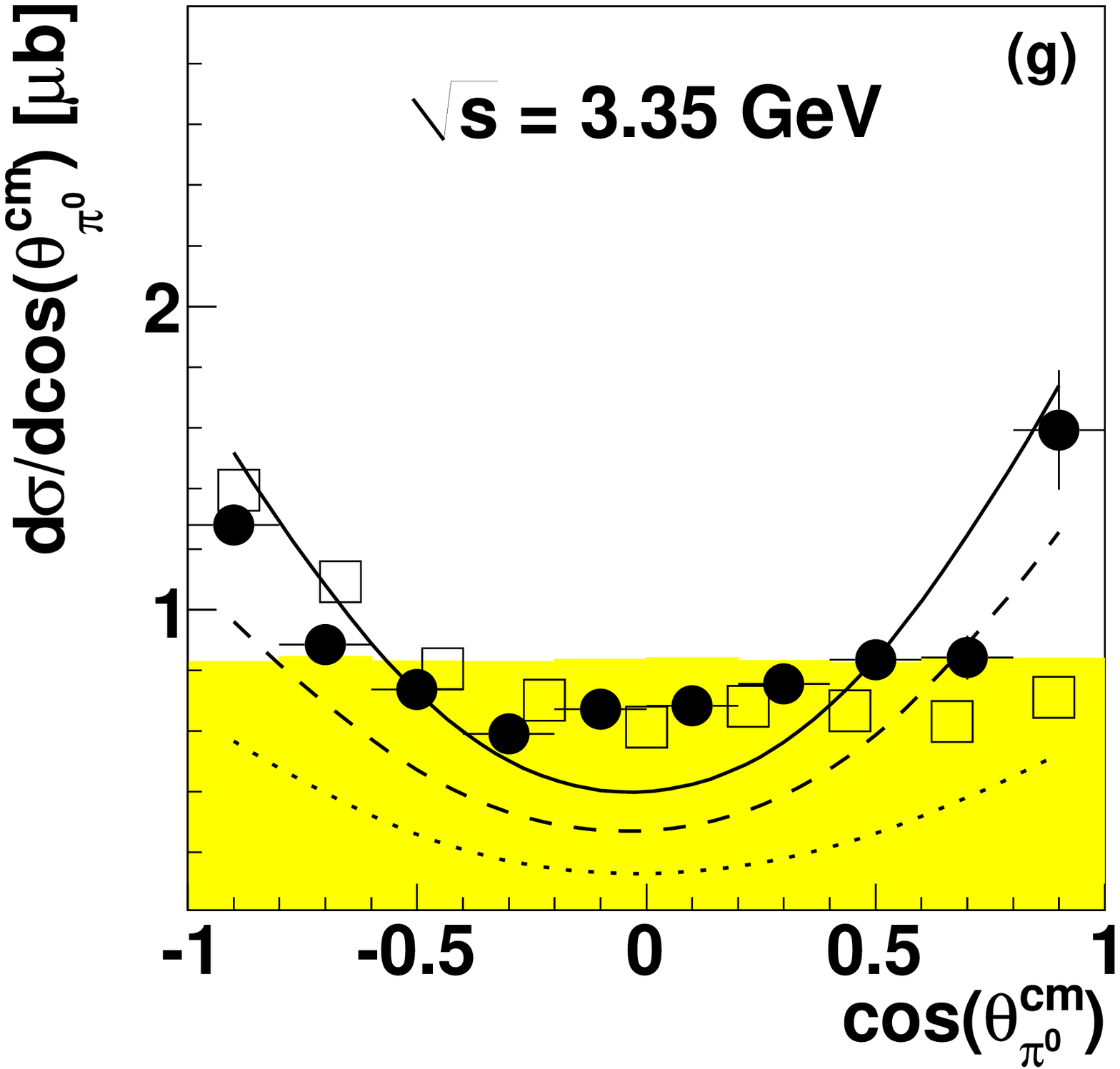}
\includegraphics[width=0.49\columnwidth]{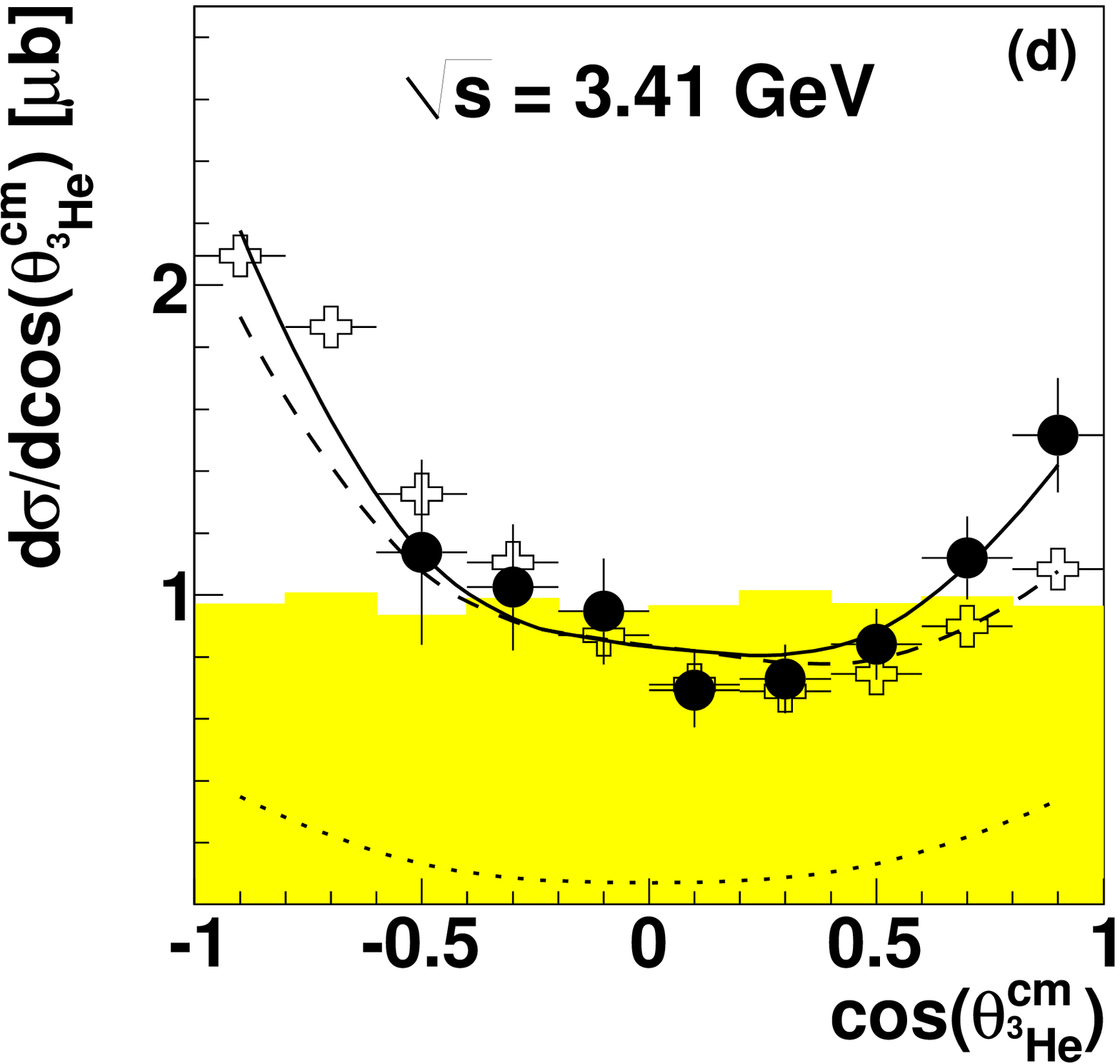}
\includegraphics[width=0.49\columnwidth]{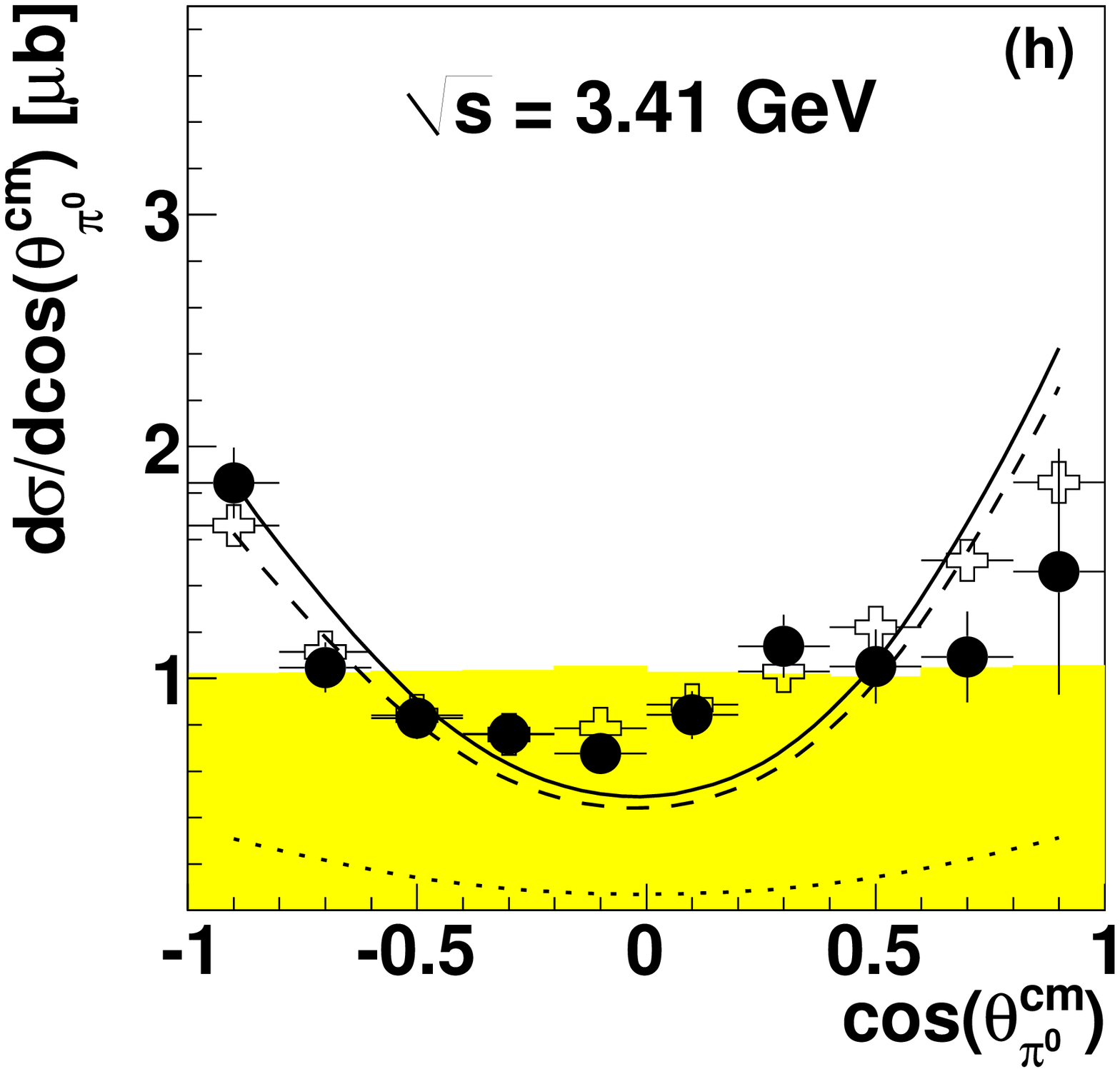}

\caption{\small Same as Fig.~5, but for the angular distributions of $^3$He
  (left) and $\pi^0$ (right) ejectiles in the center-of-mass system.
}
\end{center}
\end{figure}

For sake of completeness we note that there is also a COSY-MOMO measurement
of the $^3$He$\pi^+\pi^-$ channel at 70 MeV above threshold, {\it i.e.} at
$\sqrt s$ = 3.16 GeV \cite{Momo}. From the observed $M_{\pi^+\pi^-}$
distributions it was concluded that the produced $\pi^+\pi^-$ pair is
dominantly in relative $p$-wave, {\it i.e.} of isovector character
\cite{Momo,CW}, which is excluded in the $\pi^0\pi^0$ system discussed here. 

The energy dependence of the total cross section is consistent with some   
resonance-like structure, though we do not observe a substantial decrease
of the cross section at high energies within the measured interval. The cross
section appears to peak at a similar excess  
energy as was observed in the fusion reactions to deuterium and $^4$He.
However, as the detailed investigation of the differential
cross sections will show, the $d^*$ resonance at $\sqrt s$ = 2.37 GeV + $m_N$
shows up in the 
total cross section only as a shoulder within the ascending slope. In
marked difference to the double-pionic fusions to $d$ and $^4$He the main
contribution to the total cross section in the $^3$He case does not originate
from the $d^*$ resonance, but from the conventional $t$-channel $\Delta\Delta$
process, which has a large isovector contribution. This process peaks at around
$2m_\Delta + m_N$ and has a width of about $2\Gamma_\Delta$ \cite{isoabc,FK}.

Next we discuss the differential cross sections, which are shown in Figs.~4 -
6 and which completely describe the 3-body reaction. 
The shape of all differential distributions remains rather stable over the 
region of the $d^*$ resonance structure, however, starts to change significantly
towards the high-energy end of the measured region, where the $t$-channel
$\Delta\Delta$ process becomes dominant.

Fig.~4 shows the variation of the Dalitz plots for the invariant masses
squared  $M_{^3He\pi^0}^2$ versus $M_{\pi^0\pi^0}^2$ over the measured
resonance region. The Dalitz plots are similar to those obtained in the basic
reaction. They exhibit an enhancement in horizontal direction, in the region
of the $\Delta$ excitation, as it prominently shows up in the $M_{^3He\pi^0}$
spectra displayed in Fig.~5. This feature is consistent with the excitation of a
$\Delta\Delta$ system in the intermediate state -- as discussed for
the basic reaction \cite{prl2011,MB}.
More prominent -- and also similar to the situation in the basic
reaction -- we observe here the ABC effect as a strong enhancement at the
low-mass kinematic limit of $M_{\pi^0\pi^0}$. Consequently the
Dalitz plot is mainly populated along the $\pi\pi$ low-mass border line. 

In Fig.~5, left, the $M_{^3He\pi^0}$ distribution is shown for four selected
energies over the measured region. At all energies this distribution is far
from phase-space like (shaded areas in Fig.~5) and exhibits a clear signal
from $\Delta$ excitation.

The $M_{\pi^0\pi^0}$ distribution is shown on the right-hand side of Fig.~5 for
four different beam energies. It clearly exhibits the ABC effect at the lower
three energies. At the highest energy we see the transition to a two-hump
structure with both a low-mass enhancement and a high-mass enhancement.
The latter is the characteristic feature of the $t$-channel $\Delta\Delta$
process as predicted originally by Risser and Shuster \cite{ris} in search
for a plausible explanation of the ABC effect.

In Fig.~6 we show angular distributions at the selected energies. On the left
the angular distribution of the $^3$He ejectiles is depicted and on the right
that of the emitted $\pi^0$ particles --- both in the center-of-mass
system. Since the collision partners are not identical particles, the cm 
angular distributions do not need to be symmetric about $90^\circ$. However, in
case of a $s$-channel resonance process they have to be -- and the data appear
to be compatible with this.

The observed $^3$He angular dependence is similar to the corresponding
one in the basic reaction, though significantly more peaked near $cos\Theta
= \pm 1$ -- however, still less curved than in the double-pionic fusion to
$^4$He. 

The $\pi^0$ angular distribution resembles that for p-waves as one would
expect from the decay of $\Delta$s in the intermediate state. Note that an
intermediate $\Delta\Delta$ system shows up both in the case of $d^*$
excitation and in the case of a $t$-channel meson exchange leading to a mutual
excitation of the colliding nucleons to their first excited state, the
$\Delta$ resonance  ($t$-channel $\Delta\Delta$ process). 

Since the features, which we observe here, are very similar to those observed
for the basic double-pionic fusion reaction, we adapt the ansatz used there
for the description of the $^3$He case \cite{prl2011}. There are only two
major differences: 
\begin{itemize}
\item First, the nucleons' momenta are smeared due to their Fermi motion in
  initial and final nuclei. In particular the Fermi motion in the appreciably
  bound $^3$He nucleus leads to a sizeable smearing of the energy dependence
  in the total cross section adding nearly 30 MeV to the total width. 
\item Second, the reaction process $pd \to ^3$He$\pi^0\pi^0$ involves also the
  proton within the target deuteron, which does not participate actively in
  the formation of the $pn$ resonance, but finally forms a bound $^3$He system
  together with the $pn$ pair from the decay of the $d^*$ resonance. 
\end{itemize}

The results of this 
calculation is shown in Figs.~3, 5 - 6 by the solid lines, which provide a
reasonable description of the data. In these calculations it is assumed that
both the $d^*$ resonance and the $t$-channel $\Delta\Delta$ process happen on
the active 
$pn$ pair. The $\Delta\Delta$ process is of both isoscalar and isovector
character. From isospin coupling it follows that the latter is more than three
times as large \cite{EP}. Since only the isoscalar part of the $\Delta\Delta$
process interferes with $d^*$, the interference effect between both processes is
small. The relative size of both processes as well as the width of $d^*$
resonance has been adjusted for best reproduction of the observed
$M_{\pi^0\pi^0}$ distributions. The resulting effective $d^*$ width of 85 MeV
means that there is -- if at all -- only a small broadening due to collision
damping. It is appreciably smaller than in the $^4$He case, where the
collision broadening was about 50 MeV.  

The result of the fit to the $M_{\pi^0\pi^0}$ spectra has been scaled in
absolute height to the total cross section data in Fig.~3. We see that the
$d^*$ resonance dominates only at low energies in the strongly ascending
part of the total cross section. Thereafter the conventional $t$-channel
$\Delta\Delta$ process takes over. 

From Fig.~3 we see that the maximum cross section for $d^*$ production in the
process $pd \to d^*p \to ^3$He$\pi^0\pi^0$ is about 0.8 $\mu$b. This is a
factor of 300 less than in the basic reaction $pn \to d^* \to d\pi^0\pi^0$,
but also about a factor of two less than in the process $dd \to d^*np
\to^4$He$\pi^0\pi^0$, where there are twice as many combinatoric possibilities
to form $d^*$ in the intermediate state. This result suggests that the $d^*$
production in still heavier nuclei does not just fade away, but rather could
give sizable contributions.  

In view of the now achieved understanding of the double-pionic fusion it
appears historically rather fortunate that more than fifty years ago the
Berkeley  184-inch synchrocyclotron allowed only a maximum proton beam energy
of 743 MeV.  That way Abashian, Booth and  Crowe were in the position to just
enter the energy region, where 
the $\pi\pi$ low-mass enhancement appears to be largest -- and thus discover 
the ABC effect. Would they have had access to a beam energy of 1 GeV instead,
they would then have observed both a low-mass and a high-mass enhancement --
with the latter being the dominant one. But such a scenario was readily
explained later-on by Risser and Shuster \cite{ris} to originate naturally from
the conventional $t$-channel $\Delta\Delta$ process. Would thus the ABC
puzzle has escaped possibly its detection without providing later on the trace
to the discovery to the $d^*$ resonance?

\section{Conclusions and Outlook}

In conclusion, our data on the double-pionic fusion process to $^3$He
establish the correlation of a resonance-like energy dependence in the total
cross section with the ABC effect in very much the same way as shown before
for the 
double-pionic fusion reactions to deuterium and $^4$He. A calculation based on
the $d^*$ resonance gives a good account of the observed
distributions. The enlarged width of the resonance-structure in the total
cross section is explained by the Fermi motion of the nucleons in initial and
final nuclei, which includes also collision damping. 

That way the ABC effect in the double-pionic fusion to nuclei is
traced back to a $pn$ resonance, which obviously is strong enough to survive
even in the nuclear medium. It would be very interesting to see, whether also
in nuclei heaver than He both ABC effect and $d^*$ resonance could be
observed. Since the next heavier nuclei are not stable or do not have the
proper spin and isospin, the next suitable candidate reaction appears to be
$d ^{14}N \to ^{16}O \pi\pi$ or in inverse kinematics $^{14}N d \to ^{16}O
\pi\pi$. However, measurements of such reactions necessitate dedicated
detector setups and /or accelerators. Another great possibility to search for
ABC effect and $d^*$ resonance might be given by high-resolution measurements of
heavy ion reactions.

A trace that $d^*$ production, indeed, takes place in heavy-ion collisions
has recently been found
in connection with the so-called DLS puzzle, which denotes an unusual
enhancement in the $e^+e^-$ production at 0.3~$\lesssim M_{e^+e^-} \lesssim$
0.7 GeV initiated by neutron-proton collisions in vacuum or within
heavy-ion systems. In Ref. \cite{BC} it has been shown that a possible
solution of this puzzle is presented by accounting for $\Delta\Delta$ and $d^*$
production. 

Since a dibaryon resonance has integer spin and thus is of bosonic nature, the
survival of such a resonance in a nuclear surrounding may have an important
impact on the equation of state. Bosons are not Pauli-blocked and as such
allow for higher densities under same pressure and energy conditions. The
behavior of matter under extreme conditions is needed, {\it e.g.}, for a
better understanding of the evolution of compressed matter in the course of
heavy-ion collisions or in compact (neutron) stars --- see, {\it e.g}
Refs. \cite{kriv,AS,AF,at}.

\section{Acknowledgments}

We acknowledge valuable discussions with 
C. Hanhart, 
E. Oset and C. Wil\-kin on this issue. 
This work has been supported by BMBF, Forschungszentrum J\"ulich
(COSY-FFE), DFG, the Polish National Science Centre through the grant
No. 2011/01/B/ST2/00431 and by the Foundation for Polish Science (MPD).

\end{document}